\documentclass[%
twocolumn,
superscriptaddress,
 amsmath,amssymb,
 aps,
 prc,
]{revtex4-2}
\usepackage{comment}
\usepackage[caption=false]{subfig}
\usepackage{graphicx}
\usepackage{dcolumn}
\usepackage{bm}
\usepackage{xcolor}
\usepackage{hyperref}
\usepackage{CJKutf8}
\usepackage{lipsum}

\hypersetup{
	colorlinks=true,
	citecolor=blue,
	linkcolor=blue,
	urlcolor=blue,
}

\newcommand{\nuclide}[2]{\ensuremath{^{#1}\mathrm{#2}}}
\newcommand{\Nuclide}[2]{{^{#1}\mathrm{#2}}}

\newcommand{\kanji}[1]{\begin{CJK}{UTF8}{ipxm}(#1)\end{CJK}}

\begin{document}

\preprint{APS/123-QED}

\title{
Theoretical estimates for the synthesis of $Z=119$ superheavy nuclei with Ca, Ti, V, and Cr projectiles: effects of reaction $Q$ values and mass-model dependence
}

\author{\mbox{K.~Kawai~\kanji{川井幸亮}}}
\affiliation{Faculty of Science and Engineering, Kindai University, Higashiosaka, Osaka 577-8502, Japan}

\author{\mbox{Y.~Aritomo~\kanji{有友嘉浩}}}
\affiliation{Faculty of Science and Engineering, Kindai University, Higashiosaka, Osaka 577-8502, Japan}

\author{\mbox{K.~Nakajima~\kanji{中島滉太}}}
\affiliation{Faculty of Science and Engineering, Kindai University, Higashiosaka, Osaka 577-8502, Japan}

\author{\mbox{S.~Takagi~\kanji{髙木慎弥}}}
\affiliation{RIKEN Center for Computational Science, Kobe, Hyogo 650-0047, Japan}

\author{\mbox{N.~Nishimura~\kanji{西村信哉}}}

\affiliation{Academic Support Center, Kogakuin University, Hachioji, Tokyo 192-0015, Japan}
\affiliation{Center for Nuclear Study (CNS), The University of Tokyo, Bunkyo-ku, Tokyo 113-0033, Japan}
\affiliation{Astrophysical Big-Bang Laboratory, Pioneering Research Institute (PRI), RIKEN, Wako, Saitama 351-0198, Japan}

\date{\today}

\begin{abstract}


\begin{description}

\item[Background] The synthesis of new elements, extending the periodic table, remains one of the central challenges in modern science and requires the production of superheavy nuclei in nuclear reactions. However, the production of superheavy nuclei involves highly complex reaction mechanisms, and theoretical predictions remain subject to substantial uncertainties in nuclear-physics inputs and models.

\item[Purpose] Fusion reactions with ${}^{48}{\rm Ca}$ beams, which have been used for synthesis of $Z \le 118$ nuclei, face practical limitations for the synthesis of nuclei with $Z \ge 119$ because of the short half-lives and limited availability of suitable target nuclei. We estimate evaporation-residue cross sections $\sigma_{\rm ER}$ for the reactions ${}^{48}{\rm Ca} + {}^{254}{\rm Es}$, ${}^{50}{\rm Ti} + {}^{249}{\rm Bk}$, ${}^{51}{\rm V} + {}^{248}{\rm Cm}$, and $^{54}{\rm Cr} + {}^{243}{\rm Am}$, and examine the role of nuclear-mass-model uncertainties.

\item[Methods] We employ a hybrid framework for the three stages of the fusion reaction. The capture stage is described by the coupled-channels method, the competition between fusion and quasi-fission by a multi-dimensional Langevin approach, and the de-excitation stage by a statistical model. The $\sigma_{\rm ER}$ is evaluated by combining the corresponding capture, compound-nucleus formation, and survival probabilities.

\item[Results] Using the nuclear properties from the FRDM2012 mass model, the maximum values of $\sigma_{\rm ER}$ summed over all $x$n channels are calculated to be $233$, $206$, $33$, and $38~\mathrm{fb}$ for the ${}^{48}{\rm Ca}+{}^{254}{\rm Es}$, ${}^{50}{\rm Ti}+{}^{249}{\rm Bk}$, ${}^{51}{\rm V}+{}^{248}{\rm Cm}$, and ${}^{54}{\rm Cr}+{}^{243}{\rm Am}$ reactions, respectively. The relative relationship between the reaction $Q$ value and the Coulomb-barrier height is found to be a key factor in comparing reactions leading to the same atomic number. In particular, the relatively small $Q$ value magnitude of the ${}^{51}{\rm V}+{}^{248}{\rm Cm}$ reaction leads to a higher excitation energy and a reduced survival probability, giving the smallest $\sigma_{\rm ER}$ among the reactions considered. We also find a significant mass-model dependence on the survival probability. Using the nuclear properties predicted by several mass tables (FRDM2012, FRDM1995, WS4, and KTUY05) yields differences in the survival probability ranging from about one to several orders of magnitude. This difference mainly originates from the neutron binding energy and shell-correction energy predicted by the nuclear mass models.

\item[Conclusions] The cross sections $\sigma_{\rm ER}$ for the synthesis of $Z=119$ nuclei are governed by both the relative relationship between the reaction $Q$ value and the Coulomb-barrier height and nuclear-mass-model uncertainties that strongly affect the survival probability. Both effects must be taken into account in theoretical predictions and in the planning of future experiments.

\end{description}

\end{abstract}

\maketitle

\section{Introduction}
\label{intro}

The discovery and synthesis of new elements, that is, the extension of the periodic table, remain central challenges in modern natural science, spanning both chemistry and physics. Over the past decades, the methods used to discover new elements have shifted from chemical identification in laboratory experiments to production and identification in nuclear experiments at large accelerator facilities \cite{Oganessian_1985_summary, G.Munzenberg_1988_summary}. Today, synthesizing new elements is effectively synonymous with creating new superheavy elements, or more specifically superheavy nuclei (SHN), with atomic number $Z>103$, which requires the production of highly unstable nuclei with mass numbers of approximately 300. From the perspective of nuclear physics, this effort represents a step toward the ``Island of Stability'' \cite{MYERS19661_island_of_stability, SOBICZEWSKI1966500_island_of_stability, 2015RPPh...78c6301O} and opens an unexplored regime of nuclear structure and decay, including fission \cite{smolanczuk_PhysRevC.52.1871, SOBICZEWSKI2007292_structure, MOLLER2012, Kowal_Bf_PhysRevC.82.014303}. Progress in SHN research therefore not only extends the periodic table but also provides stringent tests of, and key insights into, modern nuclear theory.

For the production of SHN in particle-accelerator experiments, heavy-ion fusion-evaporation reactions are the primary approach. There are two principal routes: the \textit{cold fusion} reaction, employing \nuclide{208}{Pb} or \nuclide{209}{Bi} targets \cite{Oganessian_ColdFusion_1975, Hofman_RevModPhys.72.733_2000_summary, Morita_Nh_1st_2004, Morita_Nh_2nd_2007, Morita_Nh_3rd_2012}, and the \textit{hot fusion} reaction, using actinide targets \cite{Oganessian_nature114_1999, Hofmann_hotfusion_112_2007, Stavsetra_hotfusion_114_PRL103_132502_2009, Dullmann_hotfusion_114_PRL104_252701_2010, Ellison_hotfusion_114_PRL105_182701_2010, Morita_hotfusion_116_RIKEN_accel_prog_rep_2014}. Recently, the cold-fusion reactions have mainly been conducted at GSI and RIKEN, leading to the successful synthesis of elements from ${}_{107}$Bh to ${}_{113}$Nh \cite{107Bh_Muenzenberg1981, 108Hs_Münzenberg1984, 109Bh_Münzenberg1984, 110Ds_Hofmann1995, 111Bh_Hofmann1995, 112Cn_Hofmann1996, Morita_Nh_1st_2004, Morita_Nh_2nd_2007, Morita_Nh_3rd_2012}. Because the compound nucleus is formed at relatively low excitation energy, this approach benefits from an enhanced survival probability ($W$) \cite{Oganessian_ColdFusion_1975}. However, since production cross sections decrease approximately exponentially with increasing charge product of the projectile and target, cold fusion is generally considered unsuitable for synthesizing elements beyond $Z=113$ \cite{2015RPPh...78c6301O}.

In the hot-fusion approach, reactions induced by \nuclide{48}{Ca} have been widely used because they provide relatively high compound-nucleus formation probabilities. The \nuclide{48}{Ca} projectile offers advantages such as a relatively low Coulomb barrier and a large neutron-to-proton ratio associated with its doubly magic structure ($Z=20$ and $N=28$). Most of the heaviest nuclei from ${}_{114}{\rm Fl}$ to ${}_{118}{\rm Og}$ were synthesized via hot-fusion reactions with \nuclide{48}{Ca} projectiles \cite{Oganessian_Fl, Oganessian_Mc, Oganessian_Lv, Oganessian_Ts, Oganessian_Og}. However, synthesizing elements with $Z\ge119$ using \nuclide{48}{Ca} is impractical because suitable target nuclides have short half-lives and limited availability. Heavier projectiles, such as Ti, V, and Cr, have therefore attracted increasing attention because they allow the use of more stable actinide targets.

Recent experiments have demonstrated the feasibility of such heavier projectiles. At RIKEN RIBF, an experimental campaign toward $Z=119$ via the ${}^{51}\mathrm{V}+{}^{248}\mathrm{Cm}$ reaction has been carried out, including a successful measurement of the quasi-elastic barrier distribution \cite{Tanaka_119_V+Cm_RIKEN_2022}. LBNL reported the synthesis of ${}_{116}{\rm Lv}$ in the \nuclide{50}{Ti}+\nuclide{244}{Pu} reaction \cite{Gates_116_Ti_PRL133_2024}, and JINR also reported the synthesis of ${}_{116}{\rm Lv}$ using the \nuclide{50}{Ti}+\nuclide{242}{Pu} and \nuclide{54}{Cr}+\nuclide{238}{U} reactions \cite{Oganessian_Ti_Cr_EPJ_2024, Oganessian_Ti_Cr_PRC_2025, Oganessian_50Ti+242Pu_2}. These developments have intensified worldwide efforts to produce new elements beyond ${}_{118}{\rm Og}$ and to open the eighth period of the periodic table.

Although the competition to reach $Z = 119$ nuclides is intensifying~\cite{Tanaka_119_V+Cm_RIKEN_2022, Gates_116_Ti_PRL133_2024, Oganessian_Ti_Cr_EPJ_2024, Oganessian_Ti_Cr_PRC_2025, Oganessian_50Ti+242Pu_2, Khuyagbaatar_PRC102_064602_2020}, several experimental challenges remain. Even preliminary beam tests and target preparations require substantial resources. Given the extremely small production cross sections, \textit{theoretical simulations} performed in advance are essential for estimating the evaporation-residue (ER) cross section ($\sigma_{\rm ER}$) and the optimal incident energy (OIE) \cite{Hagino_2019_barrier_dist, Tanaka_PhysRevLett.124.052502_barrier_dist, Aritomo_Ohta_Langevin_NPA2004, Adamian_DNS_NPA_2000, Umar_TDHF_PhysRevC.81.064607_2010, Swiatecki_PhysRevC.71.014602_2005_FBD}. A wide variety of theoretical models has therefore been developed, including the multi-dimensional Langevin method \cite{Zagrebaev_Langevin_JPG2005, Amano_2023_PhysRevC.108.014612, Liang_Langevin_PhysRevC.87.047602, JAGANATHEN2025139302_langevin}, the dinuclear system model \cite{Nasirov_120_PhysRevC.84.044612, Jia-Xing_Cr_PRC}, the time-dependent Hartree-Fock model \cite{SIMENEL2012607_TDHF_2012, Sekizawa_TDHF+Langevin_2019_PhysRevC.99.051602, Godbey_PhysRevC.100.024610_TDHF_2019}, and the Fusion-by-Diffusion model \cite{Cap_PhysRevC.83.054602_2011_FBD, Wilczynska_120_2012_PhysRevC.86.014611_FBD, Hagino_PhysRevC.98.014607_2018_FBD}. These approaches span a broad range from macroscopic phenomenological descriptions to microscopic frameworks, and such estimates are crucial for designing efficient accelerator experiments and appropriate setups.

However, such predictions suffer from substantial uncertainties in both the underlying theoretical models~\cite{Sobiczewski_PhysRevC.90.017302, ChangGeng_DNS_MassUncertainty_PhysRevC.109.054611_2024} and the calculation methods. The SHN synthesis requires a consistent treatment of the entire reaction sequence, from the entrance channel to compound-nucleus formation and subsequent de-excitation and decay. These stages involve highly dynamical nuclear configurations and processes that are not yet fully understood from the standpoint of fundamental nuclear physics~\cite{Zagrebaev_CS_NPA}. It is therefore necessary to assess and benchmark theoretical estimates of the ER cross section and the OIE across the various approaches that have been proposed, while remaining guided by realistic experimental constraints.

In the present study, we investigate the theoretical evaluation of $\sigma_{\rm ER}$ for the synthesis of $Z = 119$ isotopes based on a hybrid framework with a dynamical description provided by the Langevin approach. We combine the coupled-channels method implemented in the CCFULL code for the capture process \cite{Hagino_CCFULL}, multi-dimensional Langevin dynamics to treat the competition between fusion and quasi-fission \cite{Zagrebaev_Langevin2, Amano_2022_PhysRevC.106.024610}, and a statistical model for the de-excitation stage \cite{Vandenbosch_sta, Reisdorf_sta_ZPA_1992, Ohta_sta}. Using this framework, we compute excitation functions of ER cross section for the reactions $\Nuclide{48}{Ca} + \Nuclide{254}{Es}$, $\Nuclide{50}{Ti} + \Nuclide{249}{Bk}$, $\Nuclide{51}{V} + \Nuclide{248}{Cm}$, and $\Nuclide{54}{Cr} + \Nuclide{243}{Am}$, which populate compound nuclei with $Z=119$. Although the Langevin-based treatment is partly phenomenological, it can reproduce existing experimental systematics with appropriately constrained physical parameters and can be extended to previously unexplored systems. We also explore the physical mechanisms that determine $\sigma_{\rm ER}$, with particular focus on the nuclear mass model, which is a major source of uncertainty in the nuclear properties entering the calculation.

This paper is organized as follows. In Section~\ref{sec:method}, we describe the theoretical framework. In Section~\ref{results}, we present the results and discuss the impact of uncertainties in nuclear mass models. In Section~\ref{summary}, we summarize the main conclusions and discuss prospects for future experiments and theoretical studies.

\section{Theoretical Framework}\label{sec:method}

\begin{figure*}[t]
    \centering
    \includegraphics[width=\linewidth]{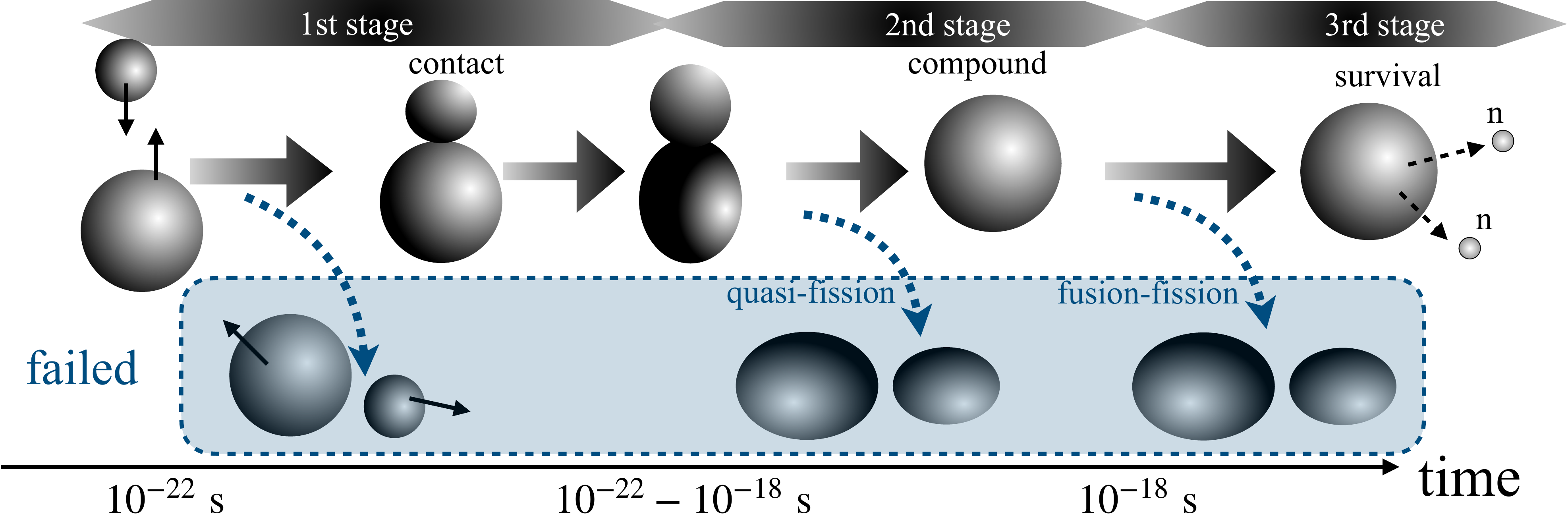}
    \caption{Schematic illustration of the calculation method and the corresponding physical stages: {\it 1st stage:} projectile--target contact, {\it 2nd stage:} the competition between compound-nucleus formation and quasi-fission, and {\it 3rd stage:} the decay of the formed compound nucleus.}
    \label{fig_calc_stage}
\end{figure*}

The synthesis of SHN can be described as a three-stage process classified by reaction timescales, as illustrated in Fig.~\ref{fig_calc_stage}. The first stage is projectile--target contact, the second stage is the competition between compound-nucleus formation and quasi-fission, and the third stage is the decay of the formed compound nucleus. In the present calculations, we employ different physical models for these stages, as described below. By combining the probabilities associated with each stage, the ER cross section $\sigma_{\mathrm{ER}}$ is evaluated as
\begin{equation}
  \sigma_{\mathrm{ER}} = \frac{\pi}{k^{2}}
     \sum_{\ell=0}^\infty (2\ell+1)\,
    T_{\ell}(E_{\mathrm{c.m.}},\ell)
     P_{\mathrm{CN}}(E^*,\ell)\,W(E^*,\ell)\;,
\end{equation}
where $k$ is the wave number of the incident flux, and $\ell$ is the orbital angular momentum in the entrance channel. $E_{\mathrm{c.m.}}$ and $E^{*}$ denote the incident energy in the center-of-mass frame and the excitation energy of the compound nucleus, respectively. The energy $E^{*}$ is given by
\begin{equation}\label{eq_energy}
E^*=E_{\mathrm{c.m.}}+Q,
\end{equation}
where $Q$ is the reaction $Q$ value calculated using the masses from the finite-range droplet model (FRDM)~\cite{MOLLER2012, Moller95}. We adopt FRDM2012 as the standard set in this study. The quantities $T_{\ell}$, $P_{\mathrm{CN}}$, and $W$ are the capture probability, the compound-nucleus formation probability, and the survival probability of the compound nucleus, respectively.

In our calculations, we adopt well-established methods for the first and third stages, namely the coupled-channels method implemented in the CCFULL code \cite{Hagino_CCFULL} and the statistical model \cite{Ohta_sta}, respectively. Since the second stage is the key part of the present work, we first summarize the treatments of the first and third stages and then describe the Langevin treatment of the fusion dynamics.

For the first stage, namely projectile--target contact, rotational excitations of heavy deformed nuclei significantly enhance the capture cross section at sub-barrier energies. We therefore apply the sudden approximation to the coupled-channels equations. Under this approximation, the capture cross section $\sigma_{\rm cap}$ is given by~\cite{Nagarajan_CC_PhysRevC.34.894, Rumin_CC_PhysRevC.63.044603, Hagino_CC_PhysRevC.69.054610}
\begin{equation}
\sigma_{\rm cap}(E_{\rm c.m.})
=\int_{0}^{1}d({\rm cos}{\theta})\sigma_{\mathrm{cap}}(E_{\mathrm{c.m.}},\theta),
\end{equation}
where $\theta$ is the orientation angle of the deformed target nucleus relative to the incident projectile direction. Assuming an axially symmetric target, $\sigma_{\rm cap}$ is calculated with an angle-dependent internuclear potential consisting of nuclear and Coulomb terms~\cite{Aritomo_CC}. The target quadrupole deformation parameter $\beta_2$, taken from the FRDM mass table~\cite{MOLLER2012, Moller95}, is incorporated into both terms. A Woods--Saxon form is adopted for the nuclear potential. Using $T_{\ell}$, the angle-dependent $\sigma_{\rm cap}$ is written as
\begin{equation}
    \sigma_{\mathrm{cap}}(E_{\mathrm{c.m.}},\theta)=
\frac{\pi}{k^2} \sum_{\ell=0}^{\infty}(2\ell+1)T_{\ell}(E_{\mathrm{c.m.}},\ell,\theta).
\end{equation}

For the third stage, the survival probability $W$ of the formed compound nucleus is calculated using the statistical model \cite{Vandenbosch_sta, Reisdorf_sta_ZPA_1992, Ohta_sta}. Using $\Gamma_{\rm n}$ and $\Gamma_{\rm f}$, which denote the decay widths for neutron evaporation and fission, respectively, $W$ is given by
\begin{equation}
W=\prod_{i=1}^{N}\frac{{\Gamma_{\rm n}}^{(i)}}{{\Gamma_{\rm n}}^{(i)} + {\Gamma_{\rm f}}^{(i)}} \ ,
\end{equation}
where $i$ labels each neutron-evaporation step, and $N$ is the total number of evaporated neutrons before the nucleus reaches an excitation energy below both the particle-emission threshold and the fission barrier.

The ratio of $\Gamma_{\rm n}$ to $\Gamma_{\rm f}$ is given by \cite{Vandenbosch_sta, Ohta_sta}
\begin{align}
    \frac{\Gamma_{\rm n}}{\Gamma_{\rm f}}
    =&\frac{4A^{2/3}a_{\rm f}(E^*-B_{\rm n})}{K_0 a_{\rm n} \left[ 2\sqrt{a_{\rm f}(E^*-B_{\rm f})} - 1 \right]}
    \cdot \frac{k_{\textrm{coll}}(\textrm{ground)}}{k_{\textrm{coll}}(\textrm{saddle}) \cdot K \cdot S} \\ \nonumber
    &\times \textrm{exp} \left[ 2\sqrt{a_{\rm n}\big(E^*-B_{\rm n}\big)}-2\sqrt{a_{\rm f}\big(E^*-B_{\rm f}\big)} \right]
\end{align}
where $A$ is the mass number of the compound nucleus and $K_0=\hbar^2/2mr_0^2$ is set to the conventional value of 10 MeV. The quantities $B_{\rm n}$ and $B_{\rm f}$ are the neutron binding energy and fission-barrier height, respectively. We use the shell-correction energy $V_{\textrm{shell}}$ predicted by several mass tables \cite{MOLLER2012, Moller95, WANG2014215_ws4, Koura_KTUY} as the fission-barrier height $B_{\rm f}$. The neutron binding energy $B_{\rm n}$ is calculated from the mass excesses in the mass tables \cite{MOLLER2012, Moller95, WANG2014215_ws4, Koura_KTUY} as
\begin{equation}
    B_{\rm n} = \Delta(Z,N-1)-\Delta(Z,N)+\Delta{n},
\end{equation}
where $\Delta(Z,N)$ is the mass excess of a nucleus $(Z,N)$ and $\Delta{\rm n}$ is the mass excess of the neutron, $\Delta{\rm n}=8.071$ MeV.

\begin{figure}[t]
    \centering
    \includegraphics[width=1.0\linewidth]{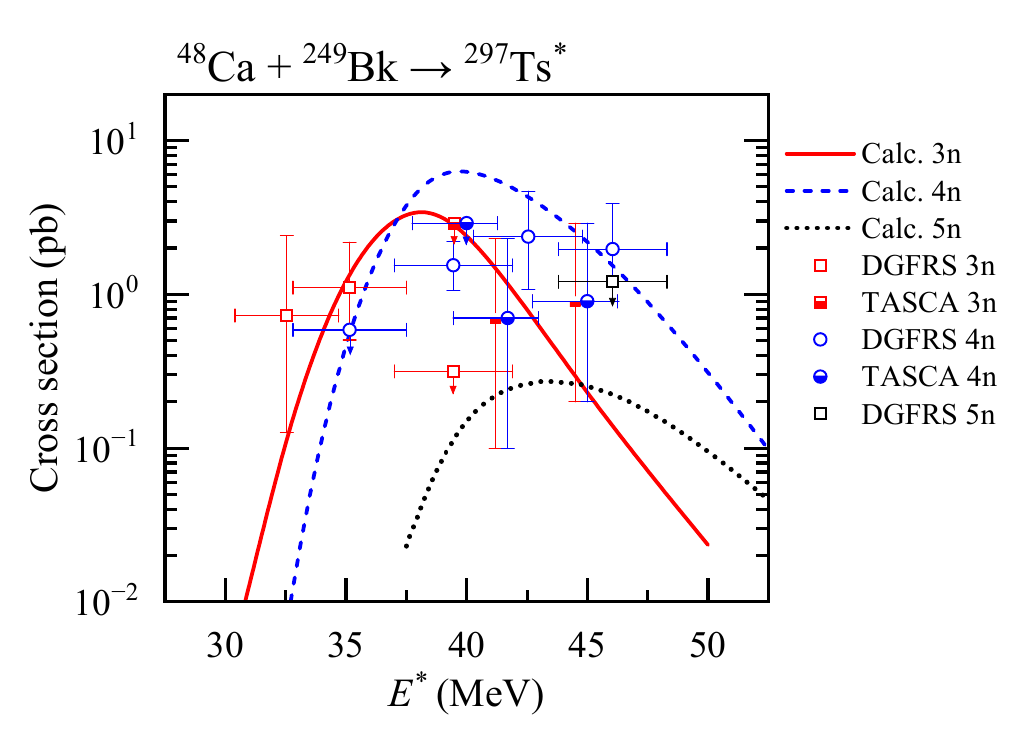}
    \includegraphics[width=1.0\linewidth]{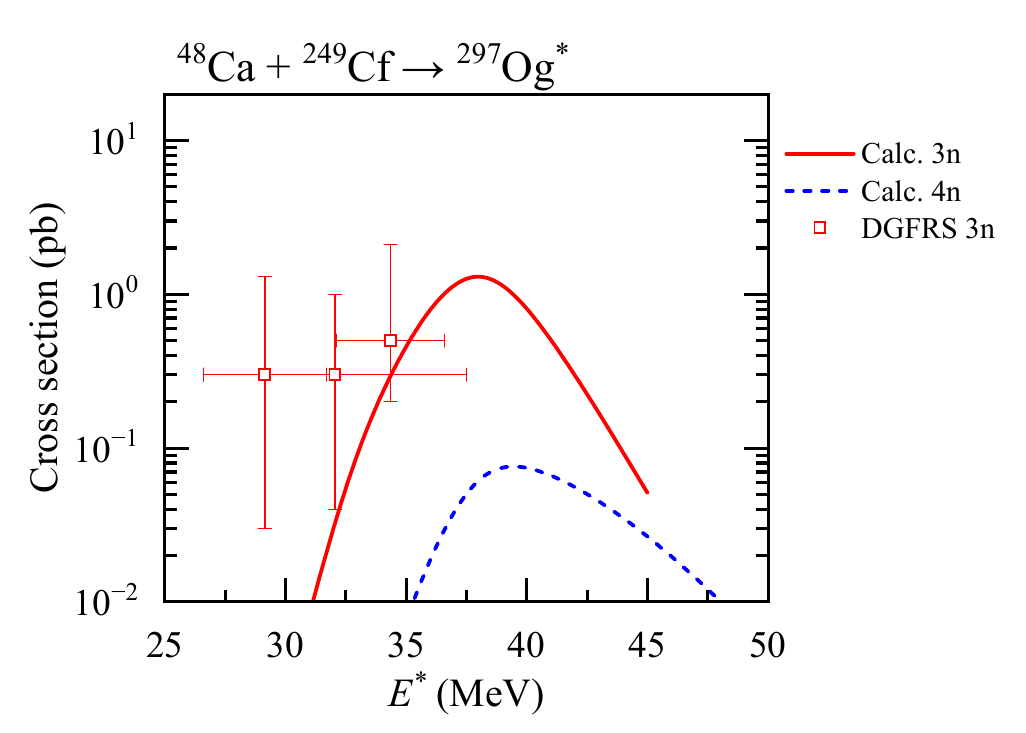}
    \caption{Excitation functions of $\sigma_{\rm ER}$ calculated with the FRDM2012 mass model for the reactions $\Nuclide{48}{Ca}+\Nuclide{249}{Bk}$, compared with the experimental data from Refs.~\cite{Oganessian_Ts, Oganessian_exTs,Khuyagbaatar_Ca+Bk_GSI_PhysRevC.99.054306}, and $\Nuclide{48}{Ca}+\Nuclide{249}{Cf}$, compared with the data from Refs.~\cite{Oganessian_Og,Oganessian_Og2_PhysRevLett.109.162501}.}
    \label{fig_ER_expt}
\end{figure}

The level-density parameters $a_{\rm n}$ and $a_{\rm f}$, which depend on $E^*$ \cite{Ignatyuc_shell}, are given by
\begin{align}
    a_{\rm n} & = \tilde{a} ~ \bigg[ 1+ \frac {V_{\textrm{shell}}} {E^*} 
    \bigg\{ 1-\textrm{exp} \bigg( -\frac {E^*} {E_\textrm{d}} \bigg) \bigg\} \bigg], \label{level_pa}\\
    a_{\rm f} &=1.02~\tilde{a},
\end{align}
where $\tilde{a}$ is the level-density parameter given by T\=oke and \'Swiatecki \cite{Toke_leveldensity_1981}. The shell-damping energy $E_\textrm{d}$ is taken to be $20~\rm{MeV}$ (see \cite{Ignatyuc_shell}). The coefficient 1.02 is determined empirically so as to reproduce the experimental data~\cite{Oganessian_Ts, Oganessian_exTs, Khuyagbaatar_Ca+Bk_GSI_PhysRevC.99.054306, Oganessian_Og, Oganessian_Og2_PhysRevLett.109.162501}, as shown in Fig.~\ref{fig_ER_expt}. 

The quantities $k_\textrm{coll}(\textrm{ground})$ and $k_\textrm{coll}(\textrm{saddle})$ denote the collective enhancement factors at the ground state and the saddle point, respectively, depending on the $\beta_2$ deformation \cite{Hagelund_1977, Bjoernholm_4268246, Ohta_sta}. The quantity $K$ denotes the Kramers factor, which corrects the fission-decay width from a dynamical perspective~\cite{KRAMERS1940284}, and is given by
\begin{equation}
    K=\sqrt{1+\bigg( \frac{\gamma}{2\omega_{\textrm{sd}}} \bigg)^2}-\frac{\gamma}{2\omega_{\textrm{sd}}},
\end{equation}
where $\gamma$ is the friction coefficient and $\omega_\textrm{sd}$ is the parameter related to the potential curvature at the saddle point. The quantity $S$ denotes the correction factor proposed by Strutinsky~\cite{STRUTINSKY1973121} for the conventional Bohr--Wheeler fission-decay width, so as to correctly account for the difference in the number of stationary collective states between the ground state and the saddle point. The $S$ factor is given by
\begin{align}
    S=\frac{\hbar\omega_{\textrm{g.s.}}}{T}, \quad 
    T=\sqrt{\frac{E^*}{a_{\rm n}}}
\end{align}
where $\hbar\omega_{\textrm{g.s.}}$ is the potential curvature at the ground state and $T$ denotes the temperature of the compound nucleus. We adopt $\gamma/2\omega_{\textrm{sd}}=1.0$ and $\hbar\omega_{\textrm{g.s.}}=1.0$ to reproduce the experimental data~\cite{Oganessian_Ts, Oganessian_exTs,Khuyagbaatar_Ca+Bk_GSI_PhysRevC.99.054306,Oganessian_Og,Oganessian_Og2_PhysRevLett.109.162501}, as shown in Fig.~\ref{fig_ER_expt}. 

The core part of our calculations is the second stage, namely the competition between compound-nucleus formation and quasi-fission. This stage is particularly complex because it involves the full dynamics of shape evolution in the superheavy-mass region, where quasi-fission is dominant and the compound-nucleus formation probability $P_{\rm CN}$ is strongly suppressed.

To evaluate $P_{\mathrm{CN}}$, we adopt a dynamical description based on multi-dimensional Langevin equations. Nuclear deformation is described using the two-center parametrization~\cite{Maruhn_TCSM, Sato_TCSM}, in terms of the three deformation coordinates $q = \{z_0, \delta, \alpha\}$. Here, $z_0$ is the distance between the centers of the two oscillator potentials, $\delta$ is the deformation parameter of the fragments, and $\alpha$ is the mass-asymmetry parameter of the colliding nuclei,
\[
\alpha = \frac{A_1-A_2}{A_1+A_2}\ ,
\]
where $A_1$ and $A_2$ denote the mass numbers of the heavy and light nuclei, respectively~\cite{Aritomo_Ohta_Langevin_NPA2004, Zagrebaev_Langevin_JPG2005}. The deformation parameter $\delta$ is defined as
\[
\delta=\frac{3(a-b)}{2a+b}\ ,
\]
where $a$ and $b$ are the half-lengths of the ellipsoidal axes in the symmetry-axis and radial directions, respectively. For simplicity, we assume that the projectile and target fragments have the same deformation $\delta$. To reduce the computation time, we introduce the scaled coordinate
\[
z=\frac{z_0}{R_{\rm{CN}}B}\ ,
\]
where $R_{\rm{CN}}$ is the radius of the spherical compound nucleus and
\[
B=\frac{3+\delta}{3-2\delta}\ .
\]
We adopt a constant neck parameter $\varepsilon = 1$ in the present calculations to retain a contact-like shape that more realistically represents the collision of two nuclei~\cite{Aritomo_Ohta_Langevin_NPA2004}.

At the beginning of the Langevin calculation, each trajectory is initialized at the touching configuration, which depends on the orientation angle $\theta$ of the deformed target nucleus. The touching distance is defined as \cite{Aritomo_CC}
\[
z^{\rm touch}_{0}(\theta)=R_P+R_T+R_T\beta_2Y_{20}(\theta),
\]
where $R_P$ and $R_T$ are the projectile and target radii, respectively. This expression explicitly accounts for the orientation of the deformed target nucleus using the previously defined $\beta_2$.
For the static deformation of the actinide targets, this $\beta_2$ is converted to the static deformation parameter $\delta_{\rm sta}$ of the two-center shell model. The relation between $\delta_{\rm sta}$ and $\beta_2$ is
\[
\delta_{\rm sta} = \frac{3 \beta_2}{ \beta_2 + \sqrt{16\pi/5} }.
\]
To describe the collision orientation of the deformed target more realistically, we further adopt the approximation proposed in Eq.~(9) of Ref.~\cite{Saiko_PhysRevC.99.014613} and determine the effective deformation parameter at contact, $\delta_{\rm touch}(\theta)$, from $\delta_{\rm sta}$:
\[
\delta_{\rm touch}(\theta)=(1+\delta_{\rm sta}) \left[ \delta_{\rm sta}(2+\delta_{\rm sta})\sin^2 \theta+1 \right]^{-\frac{3}{4}}-1.
\]
The initial mass asymmetry at contact is given by
\[
\alpha_{\rm touch} = \frac{A_T-A_P}{A_T+A_P}\ ,
\]
where $A_T$ and $A_P$ denote the mass numbers of the target and projectile nuclei, respectively.

The temperature-dependent adiabatic potential energy $V$ is defined as
\begin{align}
    V(q,L,T)&=V_{\mathrm{LD}}(q)+\frac{\hbar^2L(L+1)}{2I(q)}+V_{\mathrm{SH}}(q,T),\\
    V_{\mathrm{LD}}(q)&=E_\mathrm{S}(q)+E_\mathrm{C}(q),\\
    V_{\mathrm{SH}}(q,T)&=E^0_{\mathrm{shell}}(q)\Phi(T),\\
    \Phi(T)&=\mathrm{exp}\bigg(-\frac{\tilde{a}T^2}{E_\textrm{d}}\bigg) \ ,
\end{align}
where $L$ is the total angular momentum and $T$ is the nuclear temperature. The quantities $V_{\mathrm{LD}}$ and $V_{\mathrm{SH}}$ are the finite-range liquid-drop energy and the shell-correction energy including temperature damping, respectively. The quantity $E^0_{\mathrm{shell}}$ denotes the shell-correction energy at zero temperature \cite{STRUTINSKY_shell_1967, STRUTINSKY_shell_1968}. The temperature-dependent damping factor $\Phi(T)$ is discussed in Ref.~\cite{Aritomo_phi_t}, where $\tilde{a}$ and $E_{\textrm{d}}$ are the same parameters as those defined in Eq.~(\ref{level_pa}). The quantities $E_\mathrm{S}$ and $E_\mathrm{C}$ are the generalized surface energy \cite{Krappe_Es} and Coulomb energy, respectively. The quantity $I(q)$ denotes the moment of inertia of a rigid body with deformation $q$, and the centrifugal energy associated with $L$ is also included \cite{Aritomo_Ohta_Langevin_NPA2004}. 

\begin{figure*}[t] 
    \centering
    \subfloat{
    \includegraphics[width=0.48\linewidth]{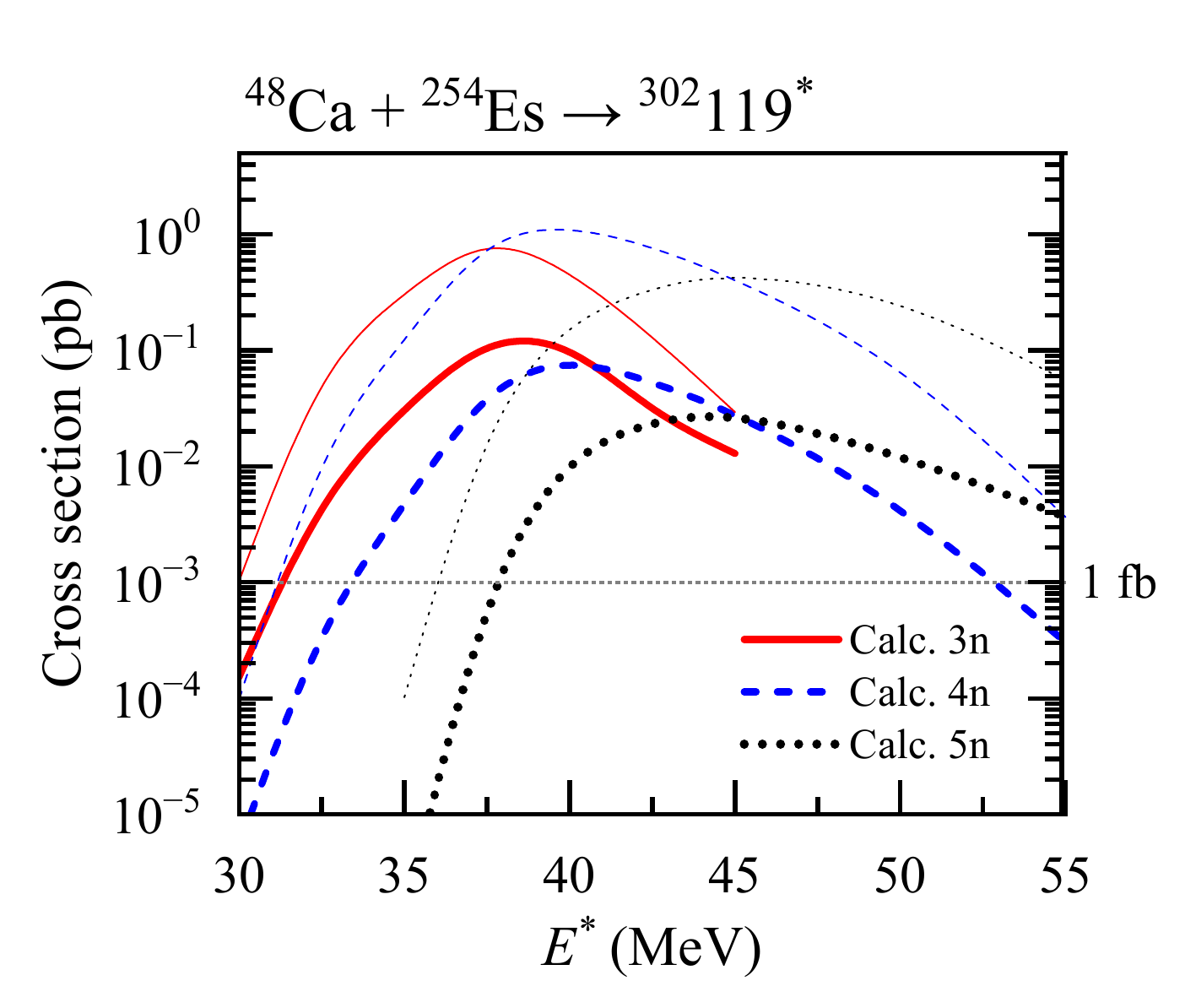}
    }
    \subfloat{
    \includegraphics[width=0.48\linewidth]{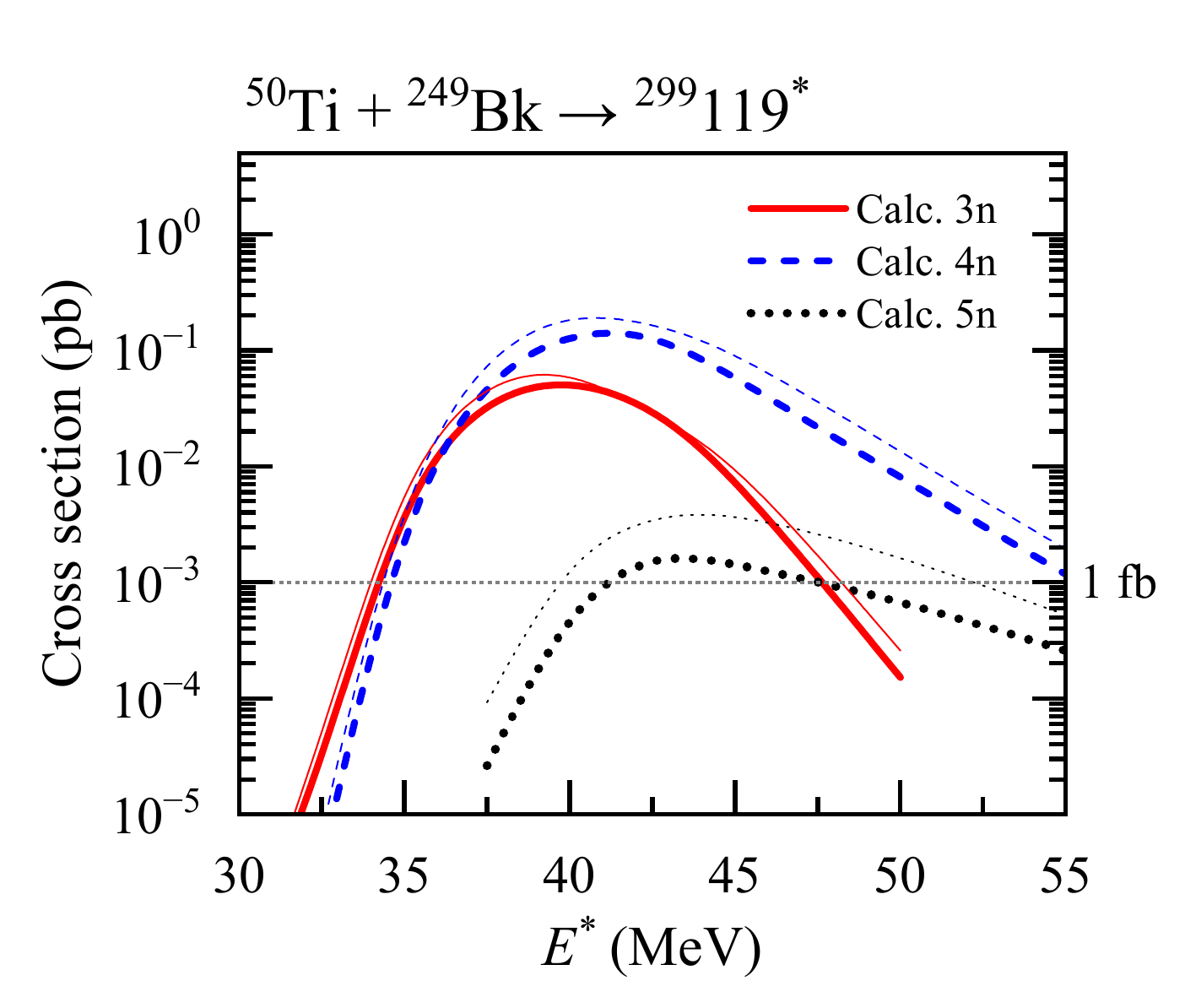}
    }

    \subfloat{
    \includegraphics[width=0.48\linewidth]{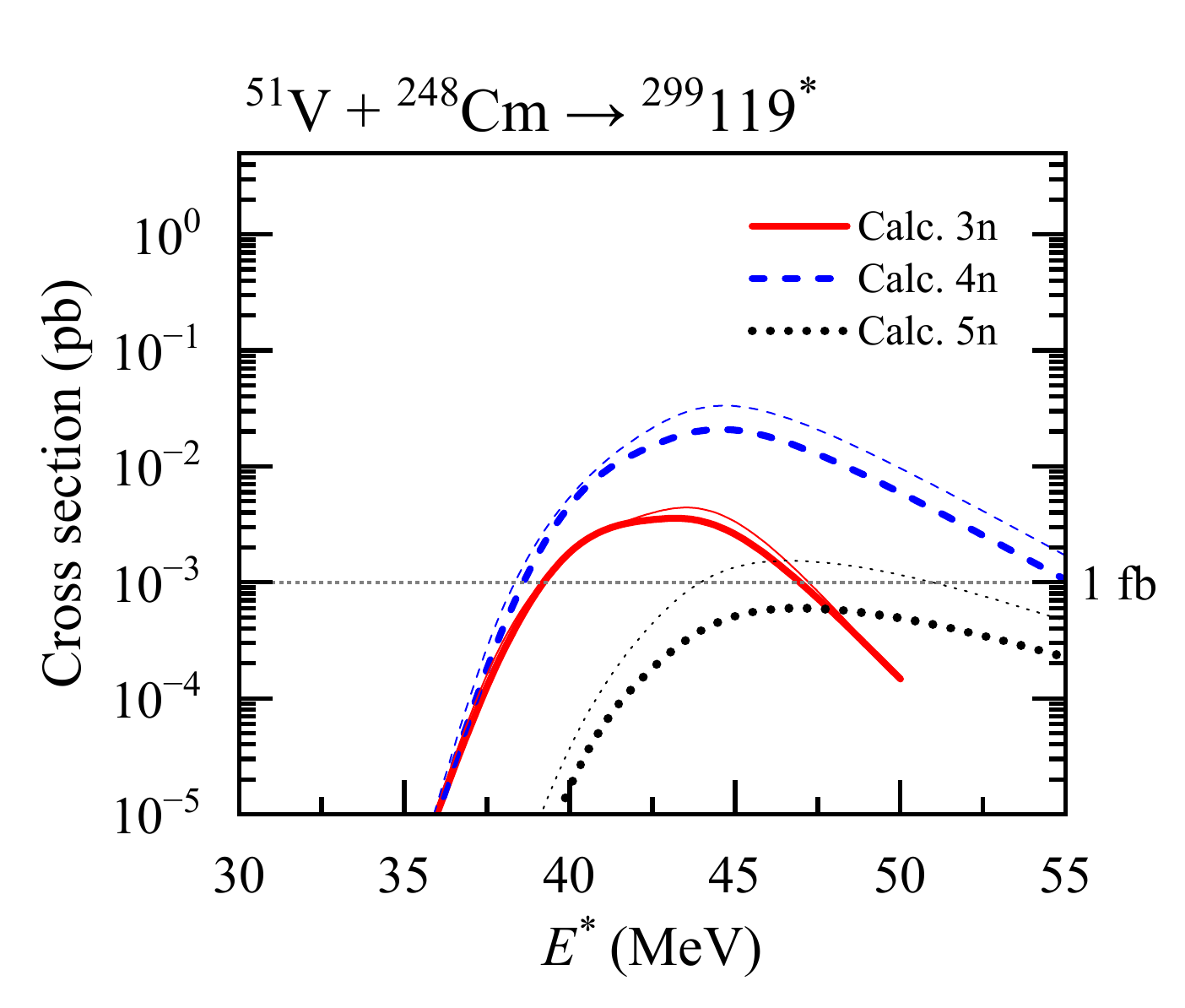}
    }
    \subfloat{
    \includegraphics[width=0.48\linewidth]{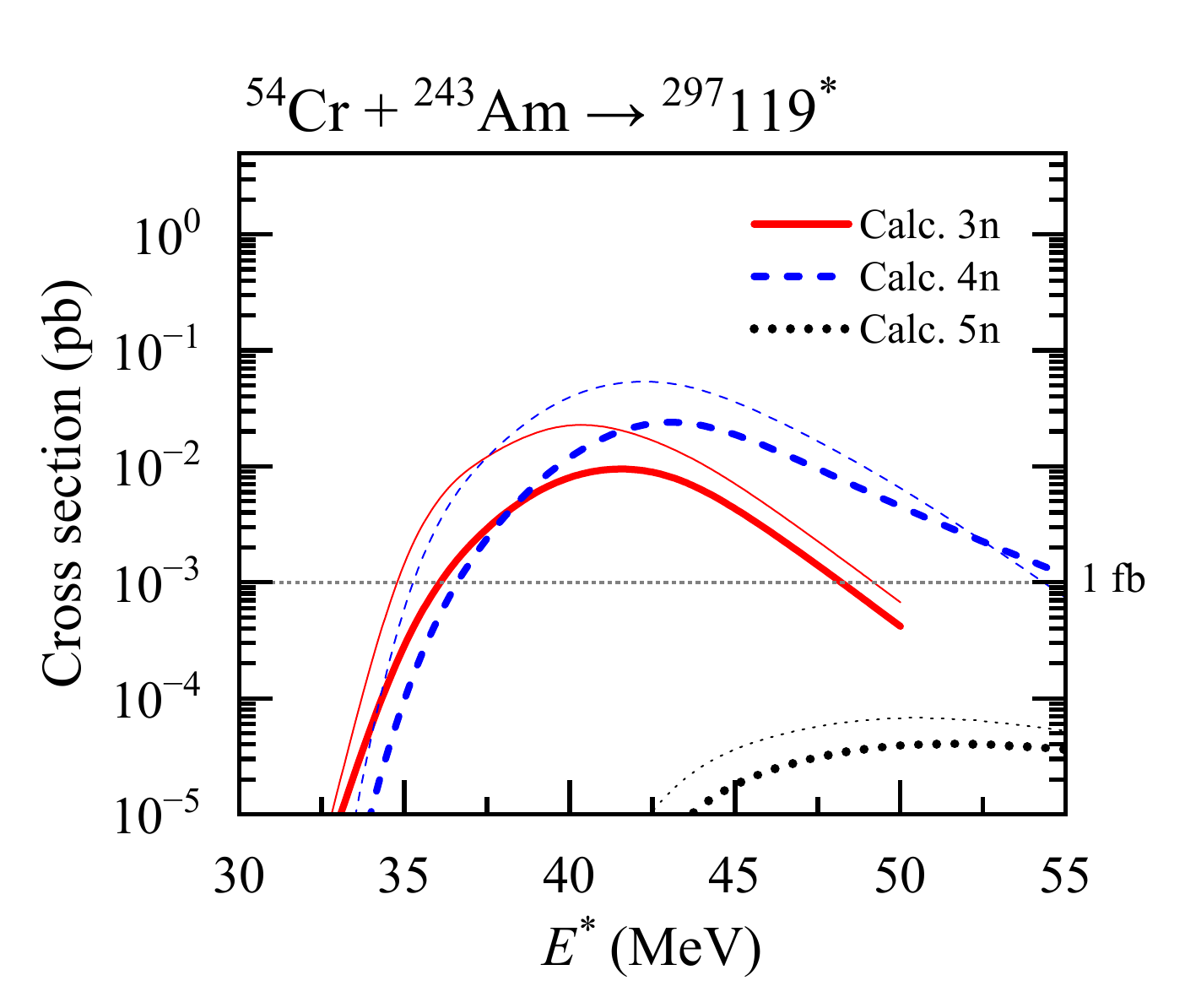}
    }
    \caption{Excitation functions of $\sigma_{\rm ER}$ for the reactions $\Nuclide{48}{Ca}+\Nuclide{254}{Es}$ (Ca--Es), $\Nuclide{50}{Ti}+\Nuclide{249}{Bk}$ (Ti--Bk), $\Nuclide{51}{V}+\Nuclide{248}{Cm}$ (V--Cm), and $\Nuclide{54}{Cr}+\Nuclide{243}{Am}$ (Cr--Am), calculated with the FRDM2012 (thick lines) and FRDM1995 (thin lines) mass models.}
    \label{fig_ER_119_2012}
\end{figure*}

Trajectories in the second stage on the potential-energy surface are calculated by solving the multi-dimensional Langevin equation \cite{Aritomo_Ohta_Langevin_NPA2004, Zagrebaev_Langevin_JPG2005, Amano_2022_PhysRevC.106.024610}:
\begin{align}    
  \frac{dq_i}{dt} =& {(m^{-1})}_{ij}p_{j}, \nonumber\\
  \frac{dp_i}{dt} =& - \frac{{\partial}V}{{\partial}q_i} - \frac{1}{2}\frac{\partial}{{\partial}q_{i}}(m^{-1})_{jk}p_{j}p_{k}\nonumber\\
  & - \gamma_{ij}{(m^{-1})}_{jk}p_{k} + g_{ij}R_{j}(t) \nonumber\\
  \frac{d {\vartheta}}{dt}=&\frac{\ell}{\mu_{R}R^{2}}, \quad\nonumber
  \frac{d \varphi_{1}}{dt}=\frac{L_{1}}{{\Im}_1}, \quad\nonumber 
  \frac{d \varphi_{2}}{dt}=\frac{L_{2}}{{\Im}_2},  \nonumber\\
  \frac{d \ell}{dt}=&-\frac{\partial V}{\partial \vartheta}-\gamma_{\mathrm{tan}}\left( \frac{\ell}{\mu_{R}R}-\frac{L_{1}}{\Im_{1}}a_{1}-\frac{L_{2}}{\Im_{2}}a_{2}\right)R \nonumber\\&+Rg_{\mathrm{tan}}R_{\mathrm{tan}}(t), \nonumber\\
  \frac{d L_{1}}{dt}=&-\frac{\partial V}{\partial \varphi_{1}}+\gamma_{\mathrm{tan}}\left( \frac{\ell}{\mu_{R}R}-\frac{L_{1}}{\Im_{1}}a_{1}-\frac{L_{2}}{\Im_{2}}a_{2}\right)a_{1} \nonumber\\&-a_{1}g_{\mathrm{tan}}R_{\mathrm{tan}}(t),  \nonumber\\
  \frac{d L_{2}}{dt}=&-\frac{\partial V}{\partial \varphi_{2}}+\gamma_{\mathrm{tan}}\left( \frac{\ell}{\mu_{R}R}-\frac{L_{1}}{\Im_{1}}a_{1}-\frac{L_{2}}{\Im_{2}}a_{2}\right)a_{2} \nonumber\\&-a_{2}g_{\mathrm{tan}}R_{\mathrm{tan}}(t)\ ,
\end{align}
where $p_i$ denotes the conjugate momentum associated with the collective coordinate $q_i$. The angle $\vartheta$ and the quantity $\ell$ denote the relative orientation of the nuclei and their relative orbital angular momentum, respectively. The angles $\varphi_1$ and $\varphi_2$ denote the rotational angles of the two nuclei in the reaction plane. Their moments of inertia and angular momenta are $\Im_{1,2}$ and $L_{1,2}$, respectively. The quantities
\[
a_{1,2}=\frac{R}{2} \pm \frac{R_1-R_2}{2}
\]
are the distances from the centers of the fragments to the midpoint between the nuclear surfaces, where $R_{1,2}$ are the nuclear radii and $R$ is the distance between the nuclear centers. The quantity $\mu_R$ is the reduced mass, and $\gamma_{\rm tan}$ is the tangential friction coefficient for the colliding nuclei~\cite{Zagrebaev_Langevin_JPG2005}. The shape-dependent collective inertia tensor $m_{ij}$ is calculated using the Werner--Wheeler approximation \cite{Davies_WW}, while the friction tensor $\gamma_{ij}$ is evaluated with the wall-and-window formula~\cite{Blocki_fric, Randrup_fric, Feldmeier_fric, Carjan_fric, Wada_fric, Asano_fric_JNRS_2006}. The random force $R_i(t)$ is assumed to be white noise, namely
\[
\langle R_i(t) \rangle =0 \quad {\rm and} \quad 
\left\langle R_i(t_1)R_j(t_2) \right\rangle =2\delta_{ij}\delta(t_1-t_2)\ .
\]
According to the Einstein relation, the strength of the random force $g_{ij}$ is given by
\[
\gamma_{ij}T=\sum_k g_{ik}g_{jk}.
\]
The temperature $T$ is related to the intrinsic energy of the composite system, $E_{\mathrm{int}}$, which is calculated at each trajectory step as
\begin{equation}
\begin{split}
     E_\mathrm{int} &= E^* - \frac{1}{2} (m^{-1})_{ij} p_i p_j -V (q, L, T=0) \\
     &= \tilde{a}T^2,
\end{split}
\end{equation}
where $E^*$ is given by Eq.~\ref{eq_energy}.

The probability $P_{\rm CN}$ is determined as
\begin{equation}
    P_{\rm{CN}}(E^*,\ell,\theta)=\frac{N_{\rm{fusion}}(E^*,\ell,\theta)}{N_{\rm{QF}}(E^*,\ell,\theta)+N_{\rm{fusion}}(E^*,\ell,\theta)},
\end{equation}
where $N_{\rm fusion}$ and $N_{\rm QF}$ denote the numbers of fusion and quasi-fission events, respectively. A trajectory is classified as fusion if it enters the fusion region defined on the potential-energy surface; otherwise, it is classified as quasi-fission. The fusion region is defined by the condition
\[
\{|\alpha|\le0.3,~\delta\le-0.5z+0.3\},
\]
which lies inside the fission saddle point~\cite{Aritomo_CC}.

Figure~\ref{fig_ER_expt} shows the excitation functions of $\sigma_{\rm ER}$ for the reactions ${}^{48}\text{Ca} + {}^{249}\text{Bk} \to {}^{297}\text{Ts}^*$ and ${}^{48}\text{Ca} + {}^{249}\text{Cf} \to {}^{297}\text{Og}^*$, calculated with this fusion-region definition. The results show good agreement with the experimental data~\cite{Oganessian_Ts, Oganessian_exTs,Khuyagbaatar_Ca+Bk_GSI_PhysRevC.99.054306,Oganessian_Og,Oganessian_Og2_PhysRevLett.109.162501}. In the next section, we discuss $\sigma_{\rm ER}$ for the synthesis of element 119 under calculation conditions similar to those adopted in these benchmark calculations.

\section{Results}\label{results}

In this section, we present the calculated $\sigma_{\rm ER}$ for the synthesis of $Z=119$ nuclei in fusion reactions induced by Ca, Ti, V, and Cr projectiles on actinide targets. We focus on the roles of the reaction $Q$ value, the Coulomb-barrier height, and nuclear-structure inputs such as masses, neutron binding energies, and fission-barrier heights. First, we examine how the interplay between the reaction $Q$ value and the Coulomb-barrier height governs $\sigma_{\rm ER}$ for several candidate reactions (Section~\ref{sec:119nuc}). We then compare results obtained with two versions of the FRDM (1995 and 2012). Subsequently, we compare the FRDM2012 with other mass models, e.g., the Weizs\"{a}cker--Skyrme (WS4) and Koura--Tachibana--Uno--Yamada (KTUY05), emphasizing the impact of mass-model uncertainties, particularly in fission-barrier heights and neutron binding energies (Section~\ref{ER_2012vs1995}).

\subsection{ER cross sections for the synthesis of $Z=119$ isotopes}\label{sec:119nuc}

\begin{table*}[t]
\caption{\label{tab_ER}
Maximum $\sigma_{\rm ER}$ values for the synthesis of $Z=119$ isotopes, summed over all $x$n channels, together with the corresponding excitation energies $E^*$ for each reaction, calculated with the FRDM2012 and FRDM1995 mass models.}
\begin{ruledtabular}
\begin{tabular}{ccccccccc}
&\multicolumn{2}{c}{$\sigma_{\rm ER}$~(fb)}
&\multicolumn{2}{c}{$E^{*}$~(MeV)}\\
Reaction & FRDM2012 & FRDM1995 & FRDM2012 & FRDM1995 \\
\colrule
${}^{48}$Ca+${}^{254}$Es &  233 & 2067 & $40.0$ & $37.5$ \\
${}^{50}$Ti+${}^{249}$Bk &  206 & 287  & $40.0$ & $40.0$ \\
${}^{51}$V+${}^{248}$Cm  &  33 & 53  & $45.0$ & $45.0$ \\
${}^{54}$Cr+${}^{243}$Am &  38 & 78  & $42.5$ & $42.5$ \\
\end{tabular}
\end{ruledtabular}
\end{table*}

\begin{figure*}[t]
    \centering
    \includegraphics[width=1.0\linewidth]{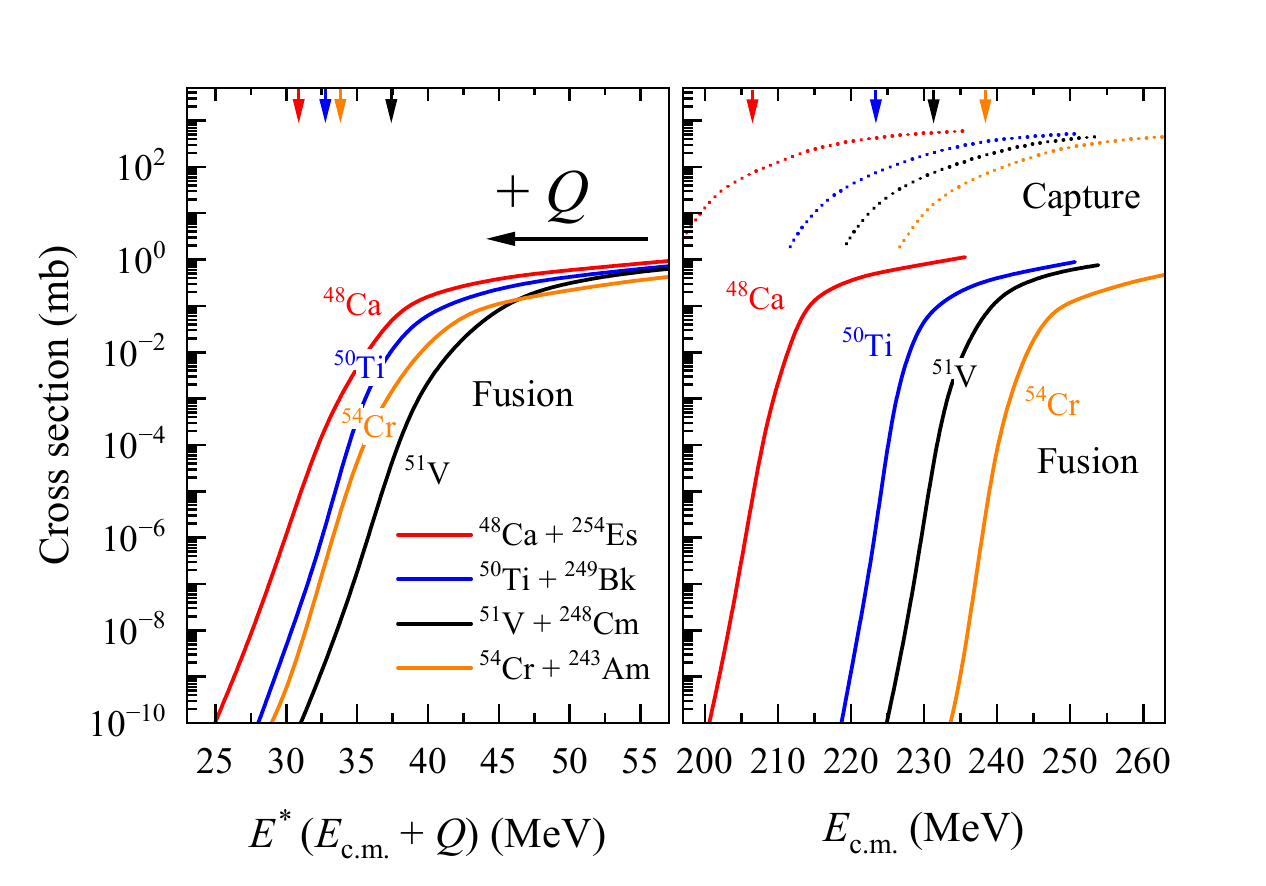}
    \caption{{\it Left panel:} Excitation functions of the fusion cross sections for the total reactions. {\it Right panel:} Capture cross sections (dotted lines) and fusion cross sections (solid lines) for each reaction, plotted as functions of the center-of-mass energy $E_{\mathrm{c.m.}}$. The arrows on the upper axis indicate the Bass barriers for the respective reactions (see Table~\ref{tab_ER} and the caption for the values).}
    \label{fig_Qval_119}
\end{figure*}

\begin{figure}[t] 
    \centering
    \includegraphics[width=0.84\linewidth]{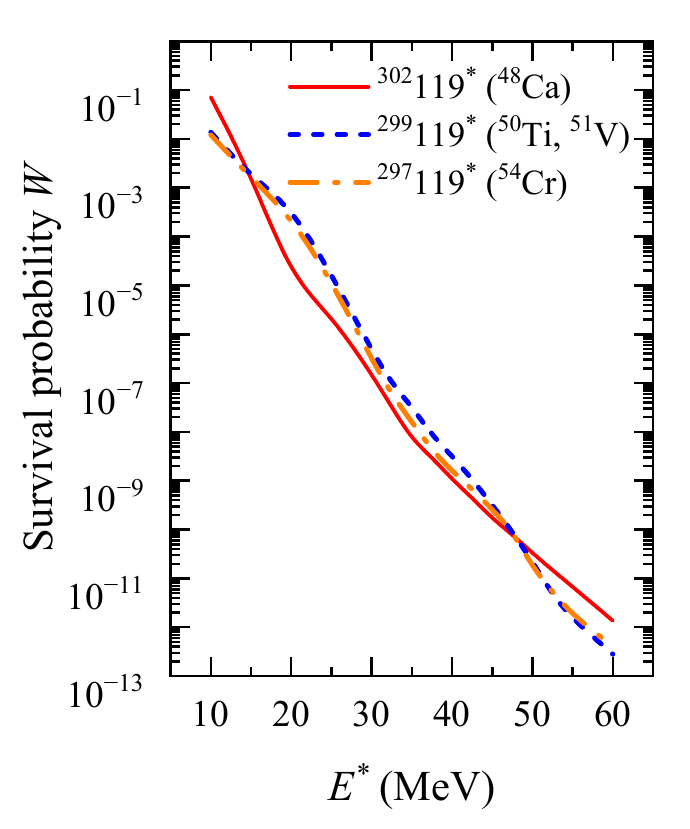} 
    \caption{Survival probabilities $W$ of $Z=119$ compound nuclei formed in different reaction systems, labeled by the target nucleus, calculated with FRDM2012.}
    \label{fig_surv2012}
\end{figure}

\begin{figure}[t] 
    \centering
    \includegraphics[width=0.9\linewidth]{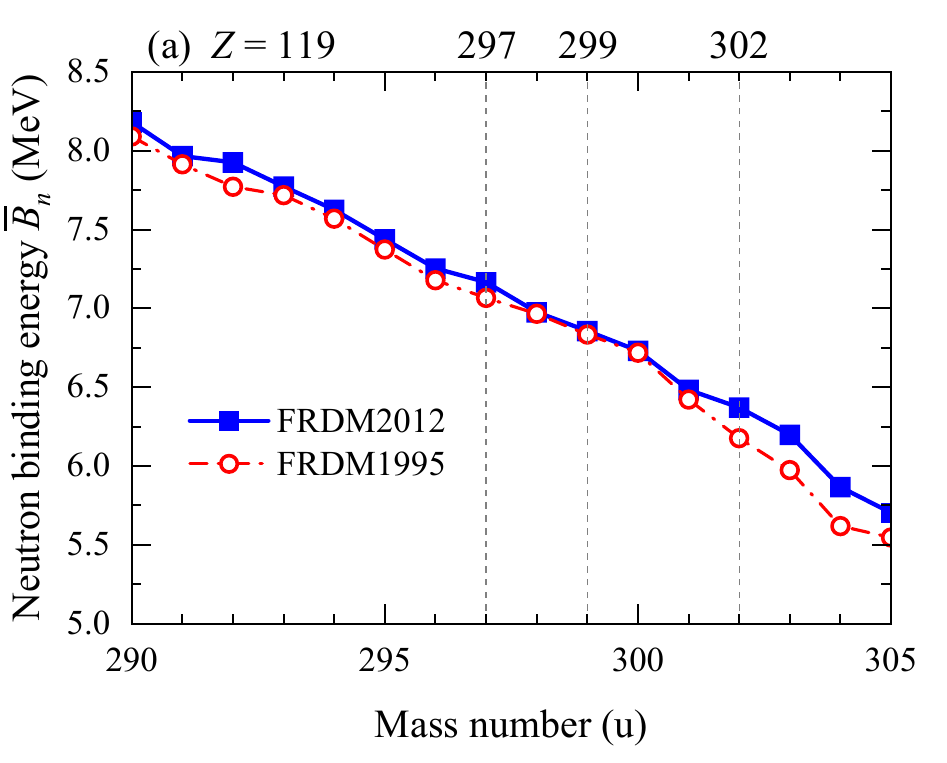} 
    \includegraphics[width=0.9\linewidth]{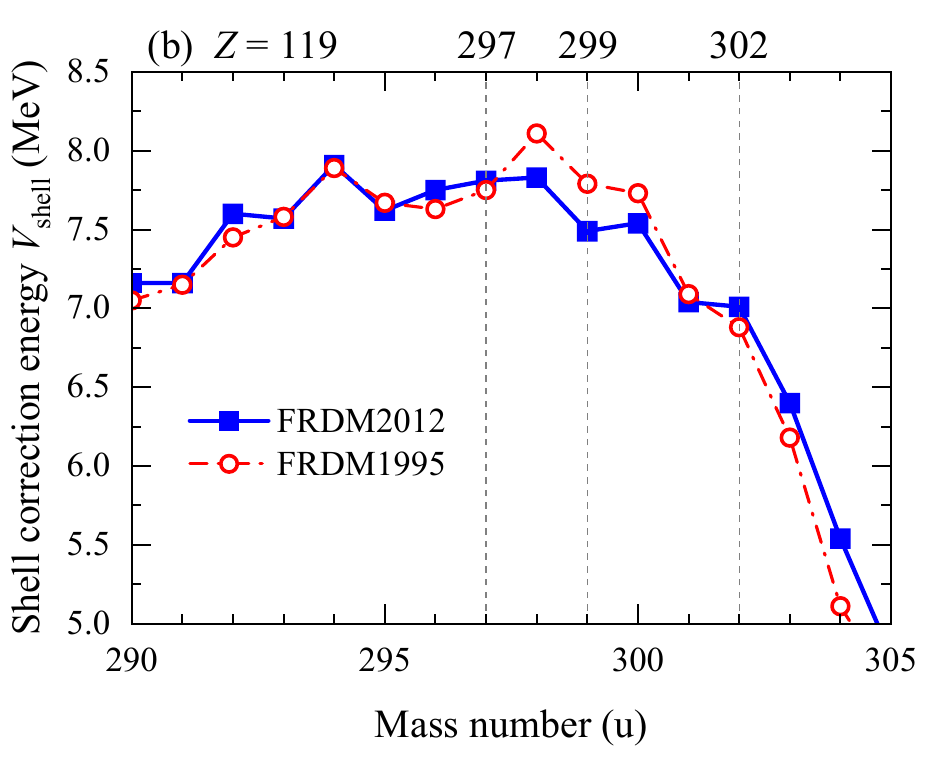} 
    \caption{(a) Neutron binding energies $\bar{B}_n$ averaged over four successive neutron emissions for isotopes with $Z=119$. (b) Shell-correction energies $V_{\mathrm{shell}}$ for isotopes with $Z=119$. The blue curves show FRDM2012 and the red curves show FRDM1995.}
    \label{fig_Vshell_2012vs1995}
\end{figure}

\begin{figure}[t] 
    \centering
    \includegraphics[width=0.84\linewidth]{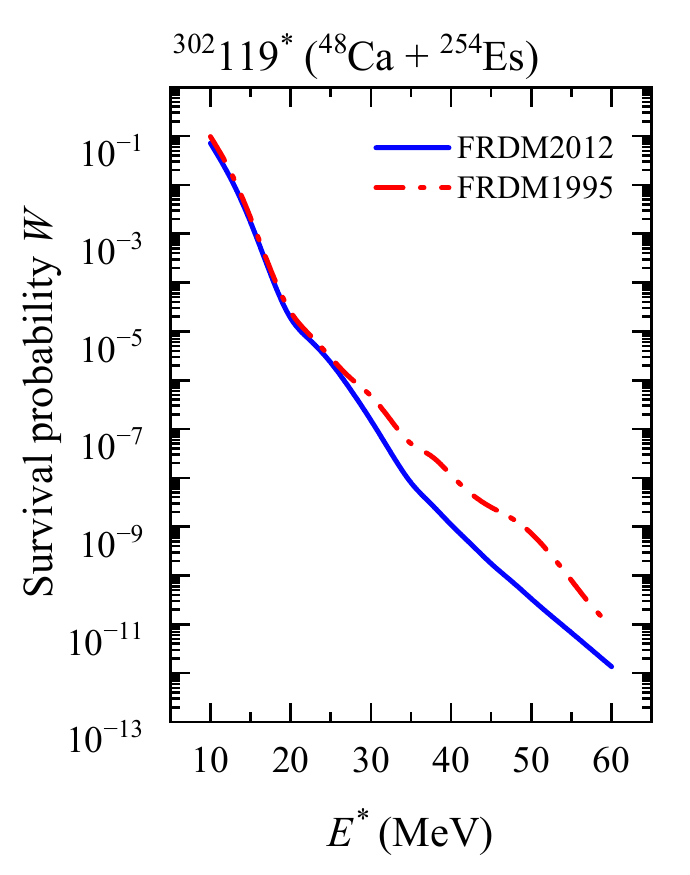}
    \caption{Survival probabilities $W$ of the compound nucleus $^{302}119$. The blue curve shows the result calculated with FRDM2012, and the red curve shows that calculated with FRDM1995.}
    \label{fig_surv_2012vs1995}
\end{figure}

\begin{table*}[t]%
\caption{\label{table_mass_119}
$Q$ values from FRDM2012 and FRDM1995, and the excitation energies $E^*$ corresponding to the center-of-mass energy $E_{\mathrm{c.m.}}$ at the Bass barrier. The excitation energy is determined by $E^* = E_{\mathrm{c.m.}} + Q$. The Bass-barrier energies are $206.49$, $223.43$, $231.31$, and $238.47~\mathrm{MeV}$ for the $\Nuclide{48}{Ca}+\Nuclide{254}{Es}$, $\Nuclide{50}{Ti}+\Nuclide{249}{Bk}$, $\Nuclide{51}{V}+\Nuclide{248}{Cm}$, and $\Nuclide{54}{Cr}+\Nuclide{243}{Am}$ reactions, respectively.}
\begin{ruledtabular}
\begin{tabular}{ccccccccc}
&\multicolumn{2}{c}{$\Nuclide{48}{Ca}+\Nuclide{254}{Es}$}
&\multicolumn{2}{c}{$\Nuclide{50}{Ti}+\Nuclide{249}{Bk}$}
&\multicolumn{2}{c}{$\Nuclide{51}{V}+\Nuclide{248}{Cm}$}
&\multicolumn{2}{c}{$\Nuclide{54}{Cr}+\Nuclide{243}{Am}$}\\
Mass table&
  $Q$ (MeV) & $E^*$ (MeV) & $Q$ (MeV) & $E^*$ (MeV) 
& $Q$ (MeV) & $E^*$ (MeV) & $Q$ (MeV) & $E^*$ (MeV) \\
\hline
FRDM2012 \cite{MOLLER2012} 
& -175.63 & 30.85 & -190.66 & 32.77 & -193.86 & 37.45 & -204.65 & 33.82\\
FRDM1995 \cite{Moller95} 
& -176.14 & 30.35 & -190.68 & 32.75 & -193.72 & 37.59 & -205.43 & 33.04\\
\end{tabular}
\end{ruledtabular}
\end{table*}

Based on the framework described in Section~\ref{sec:method}, we calculated $\sigma_{\rm ER}$ for $Z=119$ nuclei synthesis. Figure~\ref{fig_ER_119_2012} shows the excitation functions of $\sigma_{\rm ER}$ for the reactions $\Nuclide{48}{Ca} + \Nuclide{254}{Es}$ (Ca--Es), $\Nuclide{50}{Ti} + \Nuclide{249}{Bk}$ (Ti--Bk), $\Nuclide{51}{V} + \Nuclide{248}{Cm}$ (V--Cm), and $\Nuclide{54}{Cr} + \Nuclide{243}{Am}$ (Cr--Am), for which $Z_{\mathrm{proj}}Z_{\mathrm{targ}}$ values are 1980, 2134, 2208, and 2280, respectively. For FRDM2012, the maximum cross sections, summed over all $x$n evaporation channels, are 233, 206, 33, and 38~fb for the Ca--Es, Ti--Bk, V--Cm, and Cr--Am reactions, respectively, at the corresponding optimum excitation energies $E^*$ of 40, 40, 45, and 42.5 MeV. Table~\ref{tab_ER} lists the corresponding values for both mass models. In the remainder of this subsection, we discuss the FRDM2012 results as the reference case. A comparison with FRDM1995 is presented in Section~\ref{ER_2012vs1995}.

A notable feature is that the ordering of $\sigma_{\rm ER}$ is not determined simply by the charge product $Z_{\mathrm{proj}}Z_{\mathrm{targ}}$. Although Cr--Am has a larger charge product ($Z_{\mathrm{proj}}Z_{\mathrm{targ}}=2280$) than V--Cm ($Z_{\mathrm{proj}}Z_{\mathrm{targ}}=2208$), it yields a slightly larger $\sigma_{\rm ER}$. This inversion already indicates that the reaction $Q$ value and the resulting excitation energy $E^*$ play an important role through the survival probability $W$. To clarify this, we first examine the capture and fusion stages and then discuss how $W$ modifies the final ordering of $\sigma_{\rm ER}$.

\begin{figure*}[hbtp] 
    \centering
    \subfloat{
    \includegraphics[width=0.32\linewidth]{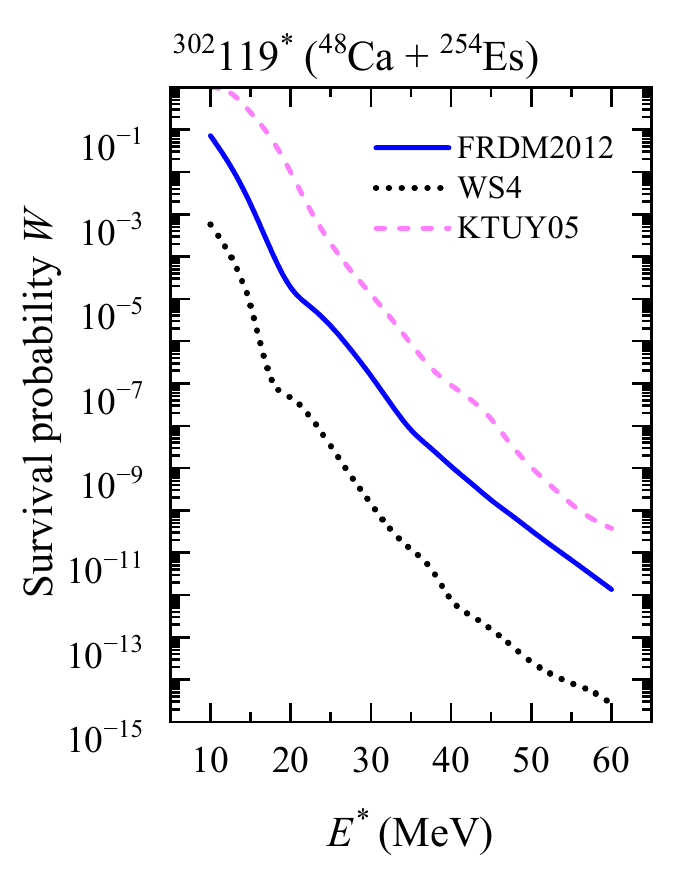}
    }
    \subfloat{
    \includegraphics[width=0.32\linewidth]{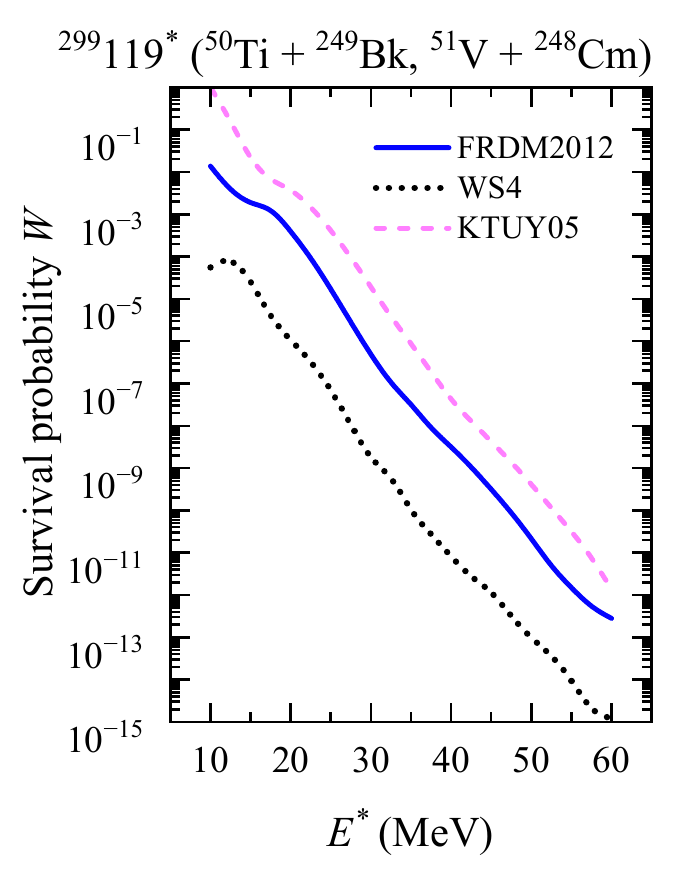}
    }
    \subfloat{
    \includegraphics[width=0.32\linewidth]{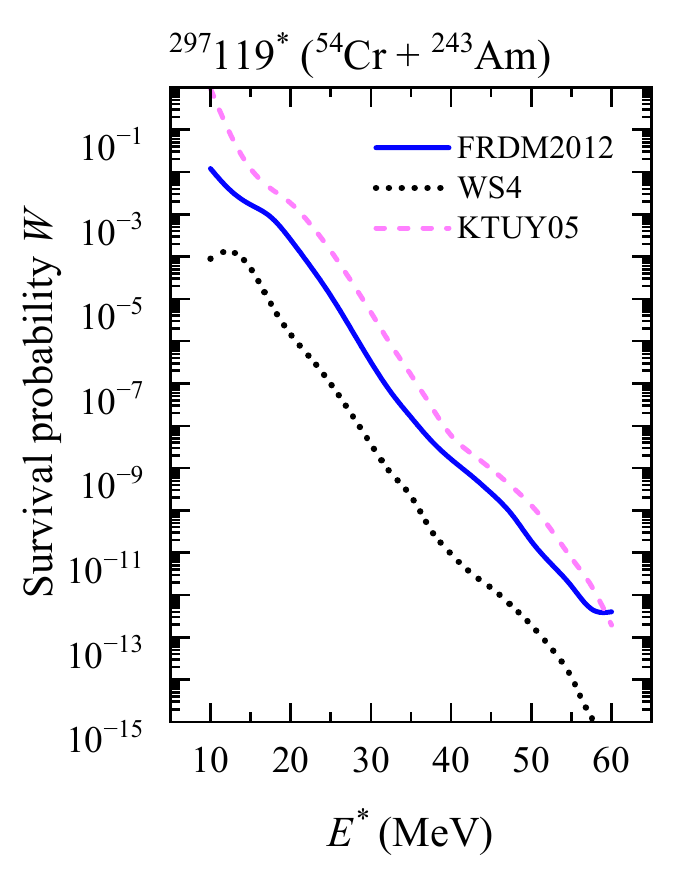}
    }
    \caption{Calculated survival probabilities $W$ of the compound nuclei $^{302}119$, $^{299}119$, and $^{297}119$. The blue solid curves show the results calculated with FRDM2012, the black dotted curves show those with WS4, and the pink dashed curves show those with KTUY05.}
    \label{fig_surv_masstable}
\end{figure*}

\begin{figure}[t]
    \centering
    \includegraphics[width=0.9\linewidth]{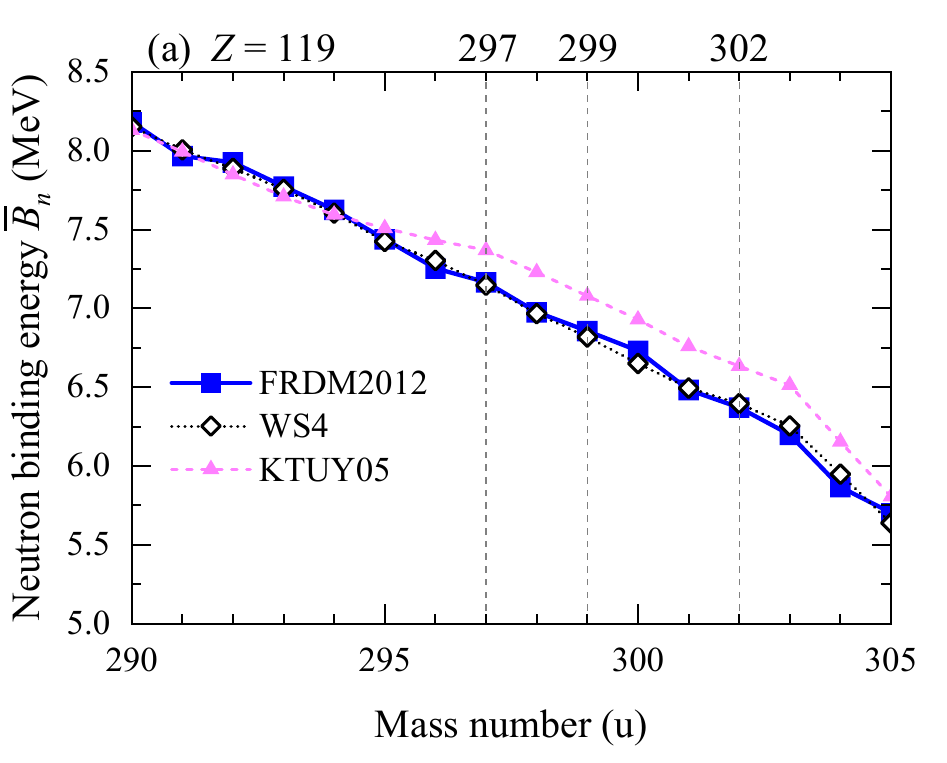} 
    \includegraphics[width=0.9\linewidth]{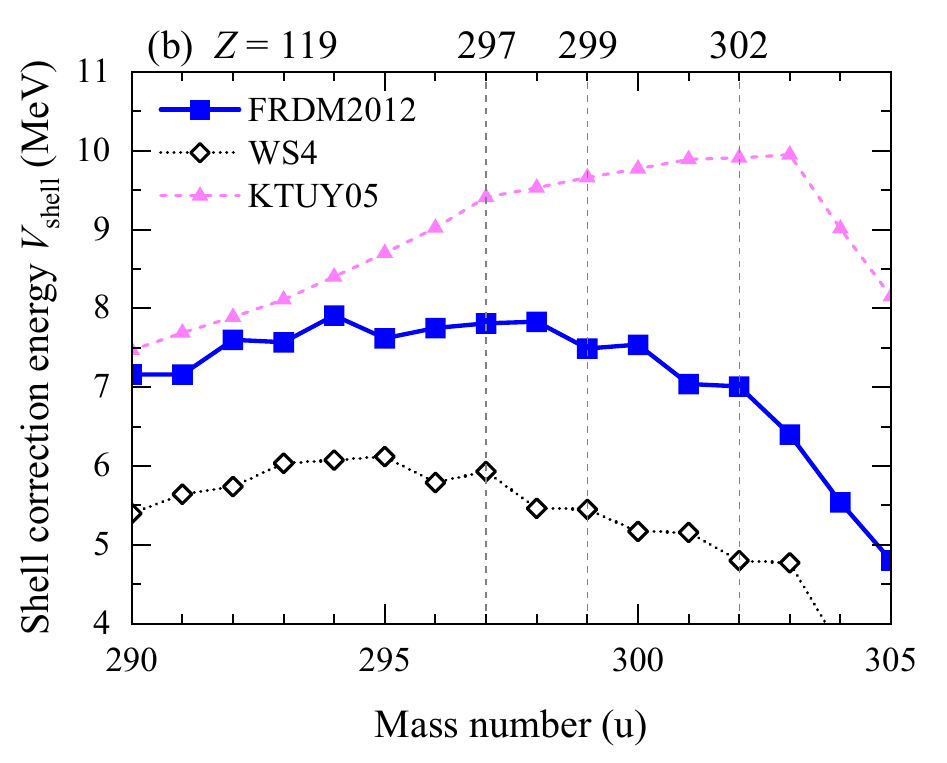} 
    \caption{(a) Neutron binding energies $\bar{B}_n$ averaged over four successive neutron emissions for isotopes with $Z=119$. (b) Shell-correction energies $V_{\mathrm{shell}}$ for isotopes with $Z=119$. The blue curves show FRDM2012, the black curves show WS4, and the pink curves show KTUY05.}
    \label{fig_nuclprop_masstable}
\end{figure}

Figure~\ref{fig_Qval_119} shows the capture and fusion cross sections corresponding to the FRDM2012 results in Fig.~\ref{fig_ER_119_2012}. The calculation sequence proceeds from the right panel to the left panel, the latter giving the final fusion cross sections as functions of $E^*$. The right panel shows the capture and fusion cross sections as functions of $E_{\mathrm{c.m.}}$, whereas the left panel shows the fusion cross sections as functions of $E^*$. The horizontal shift between the two panels reflects the relation $E^* = E_{\mathrm{c.m.}} + Q$. The $Q$ values calculated from the FRDM mass tables and the corresponding excitation energies at the Bass barrier are summarized in Table~\ref{table_mass_119}. At the Bass-barrier energy, the V--Cm reaction gives the highest $E^*$ among the reactions considered because the magnitude of its $Q$ value is relatively small compared with the Coulomb barrier.

This higher excitation energy reduces the final $\sigma_{\rm ER}$. In the right panel of Fig.~\ref{fig_Qval_119}, where the cross sections are plotted against $E_{\mathrm{c.m.}}$, reactions with lighter projectiles tend to have larger capture and fusion cross sections because of their lower Coulomb barriers. However, the relevant variable for the survival stage is $E^*$ rather than $E_{\mathrm{c.m.}}$. After converting the energy scale, the V--Cm reaction shows the smallest fusion cross section in the relevant excitation-energy region, for example around $E^*=35$--40 MeV. In other words, because its $Q$ value is relatively small in magnitude, the compound nucleus is formed at higher excitation energy once the incident energy becomes sufficient to overcome the Coulomb barrier. The fusion process therefore takes place in a region where fission more strongly suppresses $W$, and this leads to a smaller final $\sigma_{\rm ER}$.

Near the Coulomb-barrier energy, the fusion cross sections follow the descending order Ca--Es, Ti--Bk, Cr--Am, and V--Cm, as shown in Fig.~\ref{fig_Qval_119}. The final $\sigma_{\rm ER}$, however, also depends on $W$. Figure~\ref{fig_surv2012} shows the excitation functions of $W$ for $\Nuclide{302}{119}$ formed in Ca--Es, $\Nuclide{299}{119}$ formed in Ti--Bk and V--Cm, and $\Nuclide{297}{119}$ formed in Cr--Am, calculated with FRDM2012. Around $E^* \approx 40~{\rm MeV}$, which corresponds to the peak region of $\sigma_{\rm ER}$, the probabilities $W$ follow the order $\Nuclide{299}{119} > \Nuclide{297}{119} > \Nuclide{302}{119}$. This ordering is mainly determined by the fission-barrier height of the compound nucleus.

To understand this trend, Fig.~\ref{fig_Vshell_2012vs1995} shows the average neutron binding energies $\bar{B}_n$ and the shell-correction energies $V_{\rm shell}$ for $Z=119$ isotopes predicted by FRDM2012 and FRDM1995. Here we focus on the FRDM2012 results, shown by the blue curves. The isotope $\Nuclide{302}{119}$ has a smaller $V_{\rm shell}$ than the other isotopes. Although its $V_{\rm shell}$ becomes comparable to those of the others after successive neutron emissions, the initially low fission barrier strongly suppresses $W$. Nevertheless, the Ca--Es reaction still yields the largest final $\sigma_{\rm ER}$ because its fusion cross section is much larger than those of the other reactions. For $\Nuclide{297}{119}$, $W$ is lower than for $\Nuclide{299}{119}$. This difference is caused by a slightly smaller $V_{\rm shell}$ and a larger $B_{\rm n}$. The larger $B_{\rm n}$ suppresses the neutron decay width $\Gamma_{\rm n}$, thereby relatively enhancing fission.

In summary, $\sigma_{\rm ER}$ for $Z=119$ synthesis are controlled by both entrance-channel properties, such as the reaction $Q$ value and the Coulomb barrier, and compound-nucleus properties, such as $V_{\rm shell}$ and $B_{\rm n}$. The Ca--Es reaction gives the largest $\sigma_{\rm ER}$ because of its highest fusion cross section, although this advantage is partly offset by the lower $W$ of $\Nuclide{302}{119}$. The Ti--Bk reaction is also a promising candidate, supported by both a relatively high fusion cross section and the $W$ of $\Nuclide{299}{119}$. Notably, Cr--Am yields a slightly larger $\sigma_{\rm ER}$ than V--Cm despite its larger charge product. This inversion reflects the fact that the disadvantage of a higher Coulomb barrier can be compensated by a lower excitation energy and hence a larger $W$. In contrast, V--Cm is less favorable because the higher excitation energy required to overcome the Coulomb barrier suppresses $W$.

\subsection{Mass-model dependence of the ER cross section}\label{ER_2012vs1995}

As shown in the previous subsection, the calculated $\sigma_{\rm ER}$ for SHN are sensitive to several nuclear properties. Therefore, for unmeasured nuclei, the choice of nuclear mass model becomes a major source of uncertainty \cite{ChangGeng_DNS_MassUncertainty_PhysRevC.109.054611_2024, Wang_ER_uncertainty}. In this subsection, we first examine this dependence by comparing results obtained with FRDM2012 and the older FRDM1995. We then $W$ calculated from several mass models \cite{MOLLER2012,WANG2014215_ws4,Koura_KTUY} and discuss how the choice of mass model affects the predicted $\sigma_{\rm ER}$.

Figure~\ref{fig_ER_119_2012} directly compares $\sigma_{\rm ER}$ obtained with FRDM2012 and FRDM1995. For FRDM1995, the maximum $\sigma_{\rm ER}$, summed over all $x$n-evaporation channels, are $2067$, $287$, $53$, and $78~\mathrm{fb}$ for the Ca--Es, Ti--Bk, V--Cm, and Cr--Am reactions, respectively. The corresponding values for FRDM2012 are $233$, $206$, $33$, and $38~\mathrm{fb}$, as listed in Table~\ref{tab_ER}. These results demonstrate a strong mass-model dependence. In particular, for the Ca--Es reaction, the maximum $\sigma_{\rm ER}$ calculated with FRDM1995 ($2067~\mathrm{fb}$) is nearly an order of magnitude larger than that obtained with FRDM2012 ($233~\mathrm{fb}$). Although the cross sections for the other reactions are also enhanced in FRDM1995, the sensitivity is much less pronounced than in the Ca--Es case.

As listed in Table~\ref{table_mass_119}, the reaction $Q$ values differ only slightly between FRDM1995 and FRDM2012 ($\lesssim 1\%$). This indicates that the $Q$ value differences have only a minor influence on the capture and fusion stages and therefore cannot explain the large discrepancy in the Ca--Es cross section. The origin of the discrepancy must therefore lie in $W$, which is highly sensitive to the predicted nuclear properties. Figure~\ref{fig_surv_2012vs1995} compares $W$ calculated with FRDM2012 and FRDM1995. For the compound nucleus $\Nuclide{302}{119}$ formed in the Ca--Es reaction, the FRDM1995 value of $W$ is larger than the FRDM2012 value by approximately an order of magnitude in the excitation-energy region above $E^{*} = 35$~MeV. The corresponding differences in $W$ for the Ti--Bk, V--Cm, and Cr--Am reactions are much smaller, consistent with their weaker mass-model dependence in $\sigma_{\rm ER}$.

To understand the enhancement of $W$ in FRDM1995, we examine its $B_{\rm n}$ and $V_{\text{shell}}$. Previous studies have pointed out that $W$ in extremely neutron-rich systems can be enhanced by two effects: (i) low $B_{\rm n}$ and (ii) an increase in $V_{\rm shell}$ through neutron evaporation \cite{Aritomo_IoS}. Our analysis suggests that a similar mechanism operates for $\Nuclide{302}{119}$ in the FRDM1995 case. For effect (i), Fig.~\ref{fig_Vshell_2012vs1995}(a) shows that FRDM1995 predicts a slightly lower $B_{\rm n}$ than FRDM2012, which increases $\Gamma_{\rm n}$ and makes neutron evaporation more favorable. For effect (ii), Fig.~\ref{fig_Vshell_2012vs1995}(b) shows that, although the initial $V_{\rm shell}$ is small, FRDM1995 predicts a significant increase in its magnitude as neutrons are emitted. In particular, after four neutron emissions, $V_{\rm shell}$ increases by approximately 1~MeV. This recovery of the fission barrier suppresses fission in the later stages of the evaporation cascade.

As a result, in FRDM1995 the combination of low $B_{\rm n}$ and the increase in $V_{\rm shell}$ through successive neutron evaporations prevents a strong reduction of $W$ at high excitation energies ($E^{*} > 35$~MeV). Because the fusion cross section is generally large in this energy region, the enhancement of $W$ leads directly to a much larger $\sigma_{\rm ER}$. In contrast, in FRDM2012 this cooperative effect is less pronounced, resulting in a smaller $W$. This comparison suggests that the Ca--Es reaction could yield a high production rate if the nuclear properties follow the trends predicted by FRDM1995.

Thus, even different versions of the same mass model can yield substantially different nuclear properties, particularly those entering $W$. This directly propagates into uncertainties in $\sigma_{\rm ER}$ and therefore into the predicted maximum cross section and optimum excitation energy. It is therefore necessary to quantify the uncertainties in nuclear properties and carefully assess their impact on the ER-cross-section prediction.

We next examine how nuclear properties predicted by different mass formulas contribute to the uncertainty in $\sigma_{\rm ER}$. To this end, we calculate $W$ using nuclear properties from FRDM2012, WS4 \cite{WANG2014215_ws4}, and KTUY05 \cite{Koura_KTUY}, and compare the resulting differences in $W$ as an indirect measure of the uncertainty in $\sigma_{\rm ER}$.

Figure~\ref{fig_surv_masstable} compares $W$ calculated with FRDM2012, WS4, and KTUY05. For all compound nuclei considered, the use of different mass models leads to large discrepancies in $W$. Taking FRDM2012 as a reference, WS4 predicts values of $W$ that are two to three orders of magnitude smaller, whereas KTUY05 predicts values about one order of magnitude larger. To understand these differences, we compare the $B_{\rm n}$ and $V_{\rm{shell}}$ predicted by each mass model.

We first compare FRDM2012 and WS4. As shown in Fig.~\ref{fig_nuclprop_masstable}(a), there is no significant difference in $B_{\rm n}$ between the two models. However, Fig.~\ref{fig_nuclprop_masstable}(b) shows that $V_{\rm{shell}}$ in WS4 is lower than in FRDM2012 by more than 1 MeV. This smaller $V_{\rm{shell}}$ increases $\Gamma_{\rm f}$, leading to the discrepancy of two to three orders of magnitude in $W$ between FRDM2012 and WS4. Next, we compare FRDM2012 and KTUY05. As shown in Fig.~\ref{fig_nuclprop_masstable}(a), KTUY05 predicts larger $B_{\rm n}$ than FRDM2012, which suppresses $\Gamma_{\rm n}$. However, Fig.~\ref{fig_nuclprop_masstable}(b) shows that $V_{\rm{shell}}$ in KTUY05 is considerably larger than in FRDM2012. Consequently, $\Gamma_{\rm f}$ is strongly suppressed, overriding the disadvantage in $\Gamma_{\rm n}$ and ultimately enhancing $W$. This explains the discrepancy of about one order of magnitude in $W$ between FRDM2012 and KTUY05.

In summary, the differences in nuclear properties predicted by different mass models lead to severe discrepancies in $W$. Although such variations already appear between different versions of the same model, they become much larger when different mass models are compared. These variations in $W$ translate directly into significant uncertainties in the predicted $\sigma_{\rm ER}$, showing that the choice of mass model has a critical impact on the theoretical determination of optimal synthesis conditions.


\section{Summary and Conclusion} \label{summary}

In the present study, we calculated $\sigma_{\rm ER}$ for the synthesis of $Z=119$ nuclei in four reaction systems: ${}^{48}$Ca+${}^{254}$Es, ${}^{50}$Ti+${}^{249}$Bk, ${}^{51}$V+${}^{248}$Cm, and ${}^{54}$Cr+${}^{243}$Am. The cross sections $\sigma_{\rm ER}$ were evaluated within our theoretical framework, in which the capture probability $T_\ell$, the compound-nucleus formation probability $P_{\rm CN}$, and the survival probability $W$ were calculated by the coupled-channels method implemented in the CCFULL code, the Langevin approach, and the statistical model, respectively. Table~\ref{tab_ER} lists the calculated $\sigma_{\rm ER}$ values, summed over all $x$n channels, together with the corresponding excitation energies $E^{*}$, for all four reactions. Our main findings are summarized as follows:
\begin{enumerate}
\item The cross section $\sigma_{\rm ER}$ is not determined simply by the projectile--target charge product $Z_{\rm proj}Z_{\rm targ}$. Although the $\Nuclide{54}{Cr}+\Nuclide{243}{Am}$ reaction has a larger charge product ($Z_{\rm proj}Z_{\rm targ}=2280$) than the $\Nuclide{51}{V}+\Nuclide{248}{Cm}$ reaction ($Z_{\rm proj}Z_{\rm targ}=2208$), it yields a slightly larger $\sigma_{\rm ER}$. This inversion arises because the magnitude of the $Q$ value for the $\Nuclide{51}{V}+\Nuclide{248}{Cm}$ reaction is relatively small compared with its Coulomb barrier, which leads to a higher excitation energy and hence a smaller $W$.

\item The predicted $\sigma_{\rm ER}$ for the $\Nuclide{48}{Ca}+\Nuclide{254}{Es}$ reaction differs by nearly an order of magnitude between FRDM1995 and FRDM2012. This demonstrates a strong mass-model dependence of the calculated $\sigma_{\rm ER}$, even between different versions of the same model.

\item The enhancement in the FRDM1995 result originates from two cooperative effects during the evaporation process: (i) lower $B_{\rm n}$, which facilitates neutron emission, and (ii) an increase in $V_{\rm shell}$ through neutron evaporation. These effects suppress fission at high excitation energies and maintain a relatively large $W$ in that energy region.

\item Using the nuclear properties predicted by different mass models (FRDM2012, WS4, and KTUY05) yields differences in $W$ ranging from about one to several orders of magnitude. These differences arise from $B_{\rm n}$ and $V_{\rm shell}$ predicted by each mass model. The resulting variations in $W$ introduce substantial uncertainties into $\sigma_{\rm ER}$, making it difficult to determine the optimal synthesis conditions theoretically.
\end{enumerate}

Throughout this study, we analyzed the synthesis of $Z=119$ nuclei by carefully tracing each stage of the reaction sequence and by examining the underlying physical mechanisms that govern the final $\sigma_{\rm ER}$. A central conclusion is that the choice of nuclear mass model for unknown compound nuclei has a substantial impact on the predicted production cross sections. This impact is not limited to the reaction or fission $Q$ values themselves; it also appears indirectly, but often more strongly, through $W$, which is controlled by quantities such as $B_{\rm n}$ and $V_{\rm shell}$. Such effects cannot be identified from a simple comparison of $Q$ values alone, but require a stage-by-stage analysis of the full reaction process. This highlights the fact that the synthesis of superheavy nuclei is governed by a highly coupled and complex sequence of physical processes, in which entrance-channel effects, fusion dynamics, and statistical decay all play essential roles.

The present results therefore provide useful guidance for future experimental efforts to synthesize new SHN. In particular, they show that reliable theoretical predictions require not only an accurate description of capture and fusion, but also a careful assessment of mass-model uncertainties in the survival stage. The strong sensitivity of $\sigma_{\rm ER}$ to the adopted mass model, especially through $W$, implies that the theoretical determination of optimal reaction systems and excitation energies remains intrinsically uncertain unless these nuclear-property uncertainties are properly quantified. We hope that the framework and results presented here will contribute to such efforts and help guide the continued exploration of the upper end of the periodic table.
\\

The authors are grateful to K.~Hagino, K.~Nishio, M.~Ohta, and S.~Sakaguchi for fruitful discussions. Numerical computations were performed on the cluster system Kindai-VOSTOK. This work was supported by JSPS KAKENHI Grants (JP21H01087, JP24H00008, JP24K07060, JP25H01273, and JP26K00723). N.N. acknowledges support from the RIKEN Incentive Research Projects (FY2024–2025) and from the U.S. National Science Foundation under Grant No. OISE-1927130 (IReNA).


\bibliography{ref}

\clearpage

\end{document}